\documentclass[aps,prd,onecolumn,amssymb,nofootinbib]{revtex4}
\usepackage{graphicx,bm,color}
\usepackage{amsmath}
\usepackage{amssymb}
\usepackage{amsfonts}
\usepackage{epsfig}
\usepackage[T1]{fontenc}
\newcommand{\be}{\begin{equation}}
\newcommand{\ee}{\end{equation}}
\newcommand{\bea}{\begin{eqnarray}}
\newcommand{\eea}{\end{eqnarray}}
\newcommand{\beaa}{\begin{eqnarray*}}
\newcommand{\eeaa}{\end{eqnarray*}}

\allowdisplaybreaks[4]
\begin{document}
\title{Quantum cosmology in teleparallel gravity with a boundary term}
\author{H. Amiri}\email{Hamed.amiri684@gmail.com}
\author{K. Atazadeh}\email{atazadeh@azaruniv.ac.ir}
\author{H. Hadi}\email{hamedhadi1388@gmail.com}
\affiliation{Department of Physics, Azarbaijan Shahid Madani University, Tabriz, 53714-161 Iran}

\begin{abstract}
We quantize a homogeneous and isotropic universe for two models of modified teleparallel gravity, wherein an arbitrary function of the boundary term, namely $B$, is present in the action and in the other model a scalar field that is non-minimally coupled to both the torsion and boundary term.
In this regard, we study exact solutions of both the classical and quantum frameworks by utilizing the corresponding Wheeler-DeWitt (WDW) equations of the models. To correspond to the comprehensive classical and quantum levels, in the second model, we propose an appropriate initial condition for the wave packets and observe that they closely adhere to the classical trajectories and reach their peak. We quantify this correspondence using the de-Broglie Bohm interpretation of quantum mechanics. According to this proposal, the classical and Bohmian trajectories coincide when the quantum potential vanishes along the Bohmian paths.
Furthermore, we apply the de-parameterization technique to our model in the realm of the problem of time in quantum cosmological models based on the WDW equation, utilizing the global internal time denoted as $\chi$, which represents a scalar field.
\end{abstract}

\maketitle

\section{Introduction}
Current challenges in standard cosmology, such as the existence of dark energy, the accelerated expansion of the universe, the inflation paradigm, and related issues, have prompted researchers to propose and develop alternative theories of gravity. There are various approaches to constructing modified theories of gravity. One of the simplest methods involves modifying the Einstein-Hilbert action in standard general relativity (GR) or the corresponding Lagrangian by incorporating arbitrary functions of the scalars that exist within the space-time manifold. An example of such a modification is the well-known $f(R)$ modified theory of gravity, which incorporates an arbitrary function of the Ricci scalar $R$ \cite{1,11,12,13,14,2,21,22,23,24,25,3,31,32,4,41,42,5,51,52,53,54,55,56,57,58,59}.\\ 
A parallel approach to ordinary GR is the Teleparallel Equivalent of General Relativity (TEGR) where the geometrical part of the action is constructed from the torsion scalar $T$ instead of the Ricci scalar. A modification of this theory-so-called $f(T)$ gravity-was recently taken into consideration, which incorporates an arbitrary function of the torsion scalar \cite{6,7,71,72,73,74,75,76,77,78,79,710,711,712,713,714,715,716,717,718,719,720,721,722,723,724,725,726,727,728,729,730,731,732,733,734,735,736,737,739,8,9,10,111,1111,11111}.
 In the original TEGR theory and also in $f(T)$ gravity, the primary dynamical variable is the tetrad or vierbein field. This field serves as an orthonormal basis in the tangent space. The Lagrangian in this theory has a quadratic dependence on the torsion of the Weitzenb\"{o}ck connection. This connection lacks curvature and implies absolute parallelism in spacetime \cite{122}. The action of $f(T)$ gravity involves only the first derivatives of the vierbein, resulting in second-order dynamical equations. This is in contrast to $f(R)$ gravity, where the dynamical equations are fourth-order at the field equations level. Various cosmological scenarios have been explored within $f(T)$ theories of gravity, which can explain both inflationary expansions in the early universe without an inflaton field and accelerated expansion in the late times \cite{133,144}.

The problem of initial conditions poses a significant challenge in cosmological models. Unlike ordinary classical systems, which can be solved by specifying initial conditions, cosmological models lack external initial conditions that can be applied to the Einstein field equations. This is due to the absence of an external time parameter for the universe. One potential solution to this problem is the utilization of quantum cosmology, where the classical Einstein equations are replaced by a quantum Schr\"{o}dinger-like equation known as the WDW equation, along with appropriate boundary conditions \cite{a19,a191,a192}. Within the framework of quantum cosmology, various modified gravity theories have been explored, including $f(T)$ gravity \cite{darabi}, $f(R)$ gravity  \cite{a20,a201}, massive gravity \cite{a21,a211}, rainbow gravity \cite{a22}, conformally coupled scalar field gravity \cite{a23}, Horava gravity  \cite{a24,a241}, and others (refer to \cite{a25,a251,a252,a253,a254,a255,a256,shahram,gg} for further information).

In addition to the issue of initial conditions, a significant aspect concerning the WDW wave function is the ``problem of time'' \cite{z117,z118,z119,z120}. Unlike the conventional quantum theory, the wave function in quantum gravity does not depend on time. This reflects the fact that GR is a parameterized theory meaning, its action remains unchanged under time re-parameterization. Because of this problem, the canonical quantization of GR results in a constrained system, where the Hamiltonian is a combination of constraints known as Hamiltonian and momentum constraints. One possible solution to this problem is to solve the constraint equation first which allows for the derivation of a set of genuine canonical variables. These variables can then be used to construct a reduced Hamiltonian. Through this process of time re-parameterization, the equations of motion are derived from the reduced physical Hamiltonian, describing the system's evolution based on the chosen time parameter \cite{b4,b41,b42,b43,b44,b45}.

De-parameterization has emerged as a widely used technique to address the issue of time in canonical quantum gravity. Given that coordinate time is observer-dependent and lacks a corresponding operator after quantization, an alternative approach is adopted whereby a phase-space degree of freedom is chosen as a measure of change for other variables  \cite{c11,c12,c13,c14,c15,c16,c17,c18,c19}. Prominent examples of internal time include a free massless scalar field or a variable that quantifies dust. Within the scope of this study, we employ the de-parameterization to our model which is a teleparallel gravity for a quantum cosmological model, utilizing the global internal time $\chi$ which is represented by a scalar field.

Moreover, physicists have shown significant interest in the investigation of analyzing and setting the wave packets of the universe in the quantum cosmological background, as well as, their connection to the classical cosmology region. Extensive efforts have been dedicated to developing a theory that combines GR and quantum theory, known as quantum gravity, and exploring its relationship with classical aspects. Researchers have occasionally employed the semiclassical for the WDW equation, leading to the consideration of oscillatory or exponentially decaying solutions in the configuration space. These solutions correspond to allowed or forbidden regions, respectively, within the classical framework. The determination of these regions relies primarily  on the imposition of initial conditions for the wave function of the universe.
 In this work, we consider teleparallel gravity with a general form of the boundary term and we quantize this model for the special case of $f(B)=B^{2}$. In continuing to complete our study we consider the scaler-tensor teleparallel gravity in the quantum cosmology scenario to investigate the wave packets and their connection to the classical cosmology region. Additionally, we derive the Bohmian trajectory for this particular model and also we try to address the problem of time in this model.

The remainder of the paper is structured as follows: section II provides a brief overview of the teleparallel equivalence in general relativity. In section III we modified teleparallel action with a general form of the boundary term and we aim to quantize the model for the special case of $f(B)=B^{2}$.
Moving on to section IV, we delve into the examination of a modified teleparallel model, which incorporates a scalar field that is non-minimally coupled to both torsion and the boundary term. In section V, we explore the quantization and wave packets in the scalar-tensor teleparallel gravity, while section VI addresses the problem of time in this scenario. Finally, we conclude the paper in the last section.

\section{Teleparallel gravity}

The TEGR stands as an equivalent formulation of GR, referred to as the ``teleparallel'' equivalent. In contrast to the conventional approach utilizing the torsionless Levi-Civita connection, it utilizes the curvature-less Weitzenb\"{o}ck connection. The fundamental components in this framework are the four linearly independent vierbeins, which are parallel vector fields known as teleparallel.
A notable advantage of this framework is that the torsion tensor is exclusively formed from the first derivatives of the tetrad.

\begin{equation}\label{eq1}
g_{\mu\nu}=e_{\mu}^{a}e_{\nu}^{b}\eta_{ab}.
\end{equation}

The inverse tetrads are defined as

\begin{equation}\label{eq2}
e^{\mu}_{m}e^{n}_{\mu}=\delta^{n}_{m}   ,\    \   e^{\nu}_{m}e^{m}_{\mu}=\delta^{\nu}_{\mu}.\\
 \end{equation}

And the Weitzenb\"{o}ck connection is given by

\begin{equation}\label{eq3}
 W^{\ a}_{\mu\ \nu}=\partial_{\mu}e^{a}_{\ \nu}, \\
 \end{equation}

also, the torsion tensor is the antisymmetric part of the Weitzenb\"{o}ck connection which reads as

\begin{equation}\label{eq4}
 T^{a}_{\ \ \mu\nu}= W^{\ a}_{\mu\ \nu}- W^{\ a}_{\nu\ \mu}=\partial_{\mu}e^{a}_{\ \nu}-\partial_{\nu}e^{a}_{\ \mu}. \\
  \end{equation}

In addition, one can define the  torsion vector through the contraction of the torsion tensor

\begin{equation}\label{eq5}
T_{\mu}= T^{\lambda}_{\ \ \lambda\mu}.\\
 \end{equation}

To derive the field equations of teleparallel gravity, we examine the Lagrangian density below, which is subject to variation with respect to the tetrad. Unlike general relativity's Ricci scalar $R$, we incorporate the torsion scalar $T$, and the field equations of teleparallel gravity can be obtained from the following Lagrangian density by varying with respect to the tetrad \cite{Wright}

\begin{equation}\label{eq6}
{\cal L}_{T}=\frac{e}{16\pi G}S^{abc}T_{abc}, \\
 \end{equation}
 where
 \begin{equation}\label{eq7}
S^{abc}=\frac{1}{4}\left(T^{abc}-T^{bac}-T^{cab}\right)+\frac{1}{2}\left(\eta^{ac}T^{b}-\eta^{ab}T^{c}\right).\\
\end{equation}

One can define the torsion scalar $T$ as
\begin{equation}\label{eq8}
T=S^{abc}T_{abc}. \\
\end{equation}

To establish teleparallel gravity as an equivalent to GR, it is necessary to establish a connection between the Levi-Civita connection and the Weitzenb\"{o}ck connection. This connection can be expressed through the following relation.
\begin{equation}\label{eq9}
^{0}\Gamma^{\mu}_{\lambda\rho}=W^{\ \mu}_{\lambda \ \rho}-K^{\ \mu}_{\lambda \ \rho},
\end{equation}

here, $K^{\ \mu}_{\lambda \ \rho}$ is defined as contortion tensor and is given by

\begin{equation}\label{eq10}
2K^{\ \ \lambda}_{\mu \ \  \ \nu }=T^{\ \ \lambda}_{\mu \\ \nu }-T^{\ \ \lambda}_{\nu \\ \mu}+T^{\ \ \lambda}_{\mu \ \ \nu}. \
 \end{equation}

It can be observed that the contortion tensor exhibits antisymmetry in its final two indices. The Ricci scalar of the Levi-Civita connection can be expressed in terms of the Weitzenb\"{o}ck connection in the following manner.

\begin{equation}\label{eq11}
R=-T+\frac{2}{e}\partial_{\mu}(eT^{\mu}),
\end{equation}
then, one can rewrite it as

\begin{equation}\label{eq12}
R=-T+B.
\end{equation}

where $B=\frac{2}{e}\partial_{\mu}(eT^{\mu})$ is called the boundary term.
From equation (\ref{eq12}) we can see that by taking the two connections together, the torsion scalar $T$ can be written such that it is equal to the ordinary Ricci scalar, $R$, up to a boundary term. This means that we get field equations that are equivalent to the dynamical GR, also called TEGR; therefore, observations cannot be distinguished between GR and TEGR.
This boundary term is important because it contains the fourth order Ricci scalar terms, which are the boundary terms in the GR action \cite{said}.\\
Thus, we have reviewed the TEGR in a preliminary approach to establish a suitable structure for transitioning to our proposed model known as scalar-tensor teleparallel gravity, incorporating a boundary term.

\section{Teleparallel gravity with a general form of boundary term}

In this section, we consider a Friedmann-Robertson-Walker (FRW) universe in the context of teleparallel gravity with a general form of boundary term, namely, $f(B)$. Thus the proposed action can be written as\footnote{Here we set $c=\hbar=16\pi G=1$}
\begin{equation}\label{A}
{\cal S}=\int[-T+f(B)]\;e d^4x,
\end{equation}
where the quantity $e$ is defined to be the determinant of the tetrad $e^{a}_\mu$, and is equivalent to the volume element of the metric, $e =\sqrt{-g}$ and $T$ is the torsion scalar and $f(B)$ is an arbitrary function of the boundary term $B$.
We will take the standard FRW metric as

\begin{equation}\label{eq15}
ds^2=N^{2}(t)dt^2-a^2(t)\left(\frac{dr^2}{1-kr^2}+r^2(d\theta^{2}+\sin\theta^{2}d\phi^{2})\right),
\end{equation}
where $N(t)$ is the lapse function, $a(t)$ is the scale factor and $k=0,\pm 1$ plays the role of the three-dimensional space constant curvature.

To continue, we must write an effective Lagrangian for the model, and its variation with respect to the dynamical variables leads to the appropriate equations of motion. Thus, by taking the above action as representing a dynamical system in which the scale factor $a$ and boundary term  $B$ play the role of independent dynamical variables, we can rewrite the above action by considering the spatially flat FRW (\ref{eq15}) line element as

\begin{eqnarray}\label{F}
{\cal S}=\int dt {\cal L}(a,\dot{a}, B,\dot{B})=
\int dt \left\{Na^3[-T+f(B)]-\lambda\left[B+\frac{6}{N^2}\left(\frac{\ddot{a}}{a}+2\frac{\dot{a}^2}{a^2}-\frac{\dot{N}\dot{a}}{Na}\right)\right]\right\},
\end{eqnarray}
here we have inserted the definition of boundary term $B$ in terms of scale factor $a$ and its derivatives as a constraint in the Lagrangian.
 This procedure allows us to remove the second-order derivatives from action (\ref{F}).
The Lagrange multiplier $\lambda$ can be obtained by variation with respect to $ B$, that is, $\lambda = Na^3 f'(B)$, in which a prime denotes the derivative with respect to $B$.
Thus, we obtain the following point-like Lagrangian of the model
\begin{equation}\label{G}
{\cal L}=\frac{6}{N}a\dot{a}^2+\frac{6}{N}a^2\dot{a}\dot{B}f''+Na^3(f-Bf').
\end{equation}

To simplify this Lagrangian, we define the variable $\phi$ as $f'(B)=\phi$, in terms of which the Lagrangian (\ref{G}) reads
\begin{equation}\label{G1}
{\cal L}=\frac{6}{N}a\dot{a}^2+\frac{6}{N}a^2\dot{a}\dot{\phi}-Na^3V(\phi),
\end{equation}
where we define $V(\phi)=Bf'-f=B\phi-f$.

To construct the Hamiltonian of the model, the momenta conjugate to each of the above variables can be calculated from the definition
$P_{q}=\frac{\partial {\cal L}}{\partial \dot{q}}$.
 The Hamiltonian in terms of the conjugate momenta is given by
\begin{equation}\label{I}
H=\dot{a}P_a+\dot{\phi}P_{\phi}-{\cal L},
\end{equation}
Under this transformation Hamiltonian  can be written as follows
\begin{equation}\label{L}
H=N{\cal H}=N\left[\frac{P_aP_{\phi}}{6a^2}-\frac{1}{6a^3}P_{\phi}^2+a^3V(\phi)\right].
\end{equation}
We can obtain the classical dynamics that can be written by the Hamiltonian equations as follows
\begin{eqnarray}\label{M}
\begin{array}{ll}
\dot{a}=\{a,H\}=\frac{N}{6}\frac{P_{\phi}}{a^2},~~~~~~~~~~
\dot{P_a}=\{P_a,H\}=N\left[\frac{1}{3}a^{-3}P_aP_{\phi}-\frac{1}{2}\phi a^{-4}P_{\phi}^2 -3a^2V(\phi)\right],\\\\
\dot{\phi}=\{\phi,H\}=N\left[\frac{1}{6}\frac{P_a}{a^2}-\frac{1}{3}\frac{P_{\phi}}{a^3}\right],~~~~~~~~~~~~~
\dot{P_{\phi}}=\{P_{\phi},H\}=N\left[-a^3V'(\phi)\right],
\end{array}
\end{eqnarray}
In addition to the above equations, the Hamiltonian constraint equation ${\cal H}=0$ is held. Therefore, by choosing the gauge $N=a$ the classical equations of motion can be read as
\begin{eqnarray}\label{O}
\begin{array}{ll}
\dot{a}=\frac{1}{6}a^{-1}P_{\phi},~~~~~~~~~~~~~~
\dot{P_a}=\frac{1}{3}a^{-2}P_a P_{\phi}-\frac{1}{2} a^{-3}P_{\phi}^2-3a^{3}V(\phi),\\\\
\dot{\phi}=\frac{1}{6}a^{-1}P_{a}-\frac{1}{3}a^{-2}P_{\phi},~~~~~~~~~~~~~~~~~~
\dot{P_{\phi}}=-a^{4}V'(\phi),
\end{array}
\end{eqnarray}
The integrability of this system depends on how we choose the form of $f(B)$, which affects the potential $V(\phi)$ that it determines. Thus, we should focus on this choice first. However, before choosing such a function, let us consider the quantum cosmology of the model described above.

\subsection{Quantization of the model}

We will now study the quantum nature of the universe described in the model.  First, we write the WDW equation using the Hamiltonian equation (\ref{L}). Because the lapse function $N$ is used as a Lagrange multiplier in this Hamiltonian, the Hamiltonian constraint is ${\cal H}=0$. Therefore, when using the Dirac quantization procedure, the quantum states of the universe should be annihilated by the operator version of ${\cal H}$, which means that
\begin{equation}\label{P}
{\cal H}\Psi(a,\phi)=\left[\frac{P_aP_{\phi}}{6a^2}-\frac{1}{6a^3}P_{\phi}^2+a^3V(\phi)\right]\Psi(a,\phi)=0,
\end{equation}
where $\Psi(a,\phi)$ is the wave function of the universe. A remark about the reduced Hamiltonian in the above procedure
is the factor-ordering problem when one embarks on constructing a quantum mechanical operator equation. When dealing with these types of Hamiltonians in quantum physics, care must be taken when replacing the regular variables with their quantum versions. This means that when replacing a variable and its momentum with its operator versions, the order of the replacements should be considered.  So, to make sure the operator is Hermitian, we can write the operator form for equation (\ref{P}) as

\begin{eqnarray}\label{Q}
\left[\frac{1}{12}\left(a^rP_a a^s+a^s P_a a^r\right)P_{\phi}-\frac{1}{6a^3}P^{2}_{\phi}+a^3V(\phi)\right]\Psi(a,\phi)=0,
\end{eqnarray}

where the parameters $\alpha$ and $\beta$ satisfy $r+s=-2$ denote the ambiguity in the ordering of factors $a$ and $P_a$ in the first term
of (\ref{P}). With the replacement $P_a \rightarrow -i\frac{\partial}{\partial a}$ and similarly for $P_{\phi}$ the above equation reads
\begin{eqnarray}\label{R}
\left[\frac{1}{6}a^{-2}\frac{\partial^2}{\partial a \partial \phi}-\frac{1}{6}a^{-3}\frac{\partial}{\partial \phi}+\frac{1}{6}a^{-3}\frac{\partial^2}{\partial \phi^2} + a^3V(\phi)\right]\Psi(a,\phi)=0.
\end{eqnarray}

leading to
\begin{equation}\label{U}
\left[a\phi\frac{\partial^2}{\partial a \partial \phi}+\phi\frac{\partial^2}{\partial \phi^2}-\phi\frac{\partial}{\partial \phi}+
6a^6\phi V(\phi)\right]\psi(a,\phi)=0.
\end{equation}
 This equation has a combination of derivatives with respect to the variables $a$ and $\phi$, and these variables also appear together in the last part of the equation.  Under these conditions, the equation (\ref{U}) cannot be solved by splitting it into separate parts. Thus it is helpful to use these different variables instead
\begin{equation}\label{V}
u=a\sqrt{\phi},\hspace{.5cm}v=\phi.
\end{equation}
In terms of these variables equation (\ref{U}) takes the form
\begin{eqnarray}\label{W}
\left[\frac{1}{4}u^2\frac{\partial^2}{\partial
u^2}+\frac{1}{4}u\frac{\partial}{\partial u}+v\frac{\partial^2}{\partial v^2}-v\frac{\partial}{\partial
v}+6u^6v^{-2}V(v)\right]\psi(u,v)=0.
\end{eqnarray}
Unfortunately, we can not solve this equation using a formula for any potential $V(v)$ that represents the form of the function $f(B)$.
 In the next subsection, we will present a class of exact solutions for this equation with $f(B)=B^2$.

\subsection{ Quantum cosmology with the squared boundary term $f(B)=B^2$ }

By choosing $f(B)=B^2$, we have $\phi=f'(B)=2B$ and $V(\phi)=Bf'-f=B^2$. Hence, from (\ref{V}) we obtain $V(\phi)=\frac{1}{4}\phi^2=\frac{1}{4}v^2$.
Therefore, equation (\ref{W}) reduces to
\begin{equation}\label{X}
\left[\frac{1}{4}u^2\frac{\partial^2}{\partial u^2}+\frac{1}{4}u\frac{\partial}{\partial u}+v\frac{\partial^2}{\partial v^2}-
v\frac{\partial}{\partial v}+\frac{3}{2}u^6\right]\psi(u,v)=0.
\end{equation}
We can see that the variables $u$ and $v$ can be taken apart from each other and we can solve equation (\ref{X}) for each of them separately.
 Thus, in what follows we restrict ourselves to these two special cases.
In this case we separate the solutions of equation (\ref{X}) into the form $\psi(u,v)=U(u)W(v)$
leading to
\begin{eqnarray}\label{Y}
\begin{array}{ll}
\left[u^2\frac{d^2}{du^2}+u\frac{d}{du}+\left(1-\nu^2+6u^6\right)\right]U(u)=0,\\\\
\left[v\frac{d^2}{dv^2}-v\frac{d}{dv}-\frac{\nu^2-1}{4}\right]W(v)=0,
\end{array}
\end{eqnarray}
where we take $\frac{\nu^2-1}{4}$ as a separation constant.

The above equations have the following solutions in terms of Bessel and Hypergeometric functions for having well-defined functions in all ranges of variables $u$ and $v$.

\begin{eqnarray}\label{Z}
\begin{array}{ll}
U(u)=c_1J_{\frac{\sqrt{\nu^2-1}}{3}}\left(\sqrt{\frac{2}{3}}u^3\right),\\\\
W(v)=vd_{1}F_{1}[5/4 - \nu^2/4, 2,v],
\end{array}
\end{eqnarray}
where $c_1$ and $d_1$ are integration constants. Thus, the wave function of the WDW equation can be written as
\begin{equation}\label{AB}
\Psi_{\nu}(u,v)=c_{0}vJ_{\frac{\sqrt{\nu^2-1}}{3}}\left(\sqrt{\frac{2}{3}}u^3\right)F_{1}[5/4 - \nu^2/4, 2,v],
\end{equation}

where $c_0$ is a constant.
To understand the physical behavior of the wave function, we plot its square in Figure 1.
This plot shows that the wave function has a large peak near some non-zero values of $u$ and $v$, and smaller peaks after that. As the value of $u$ increases, the smaller peaks decrease. This means that the wave function can predict how the universe started from its most likely state.
The wave function has a well-defined behavior (sudden peak) near $u\sim1, v\sim 0.8$ and describes a universe, without a singularity problem, emerging out of nothing without any tunneling.
 However, when the wave function has several peaks, it could mean that different quantum states communicate with each other by tunneling.
Thus, our universe could have developed from different possible states and moved from one state to another in the past. Bearing in mind that $y=\phi=f'(B)$, this wave function predicts that the universe will assume states with larger $B$ in its late time evolution.

\begin{figure*}[ht]
  \centering
  \includegraphics[width=3in]{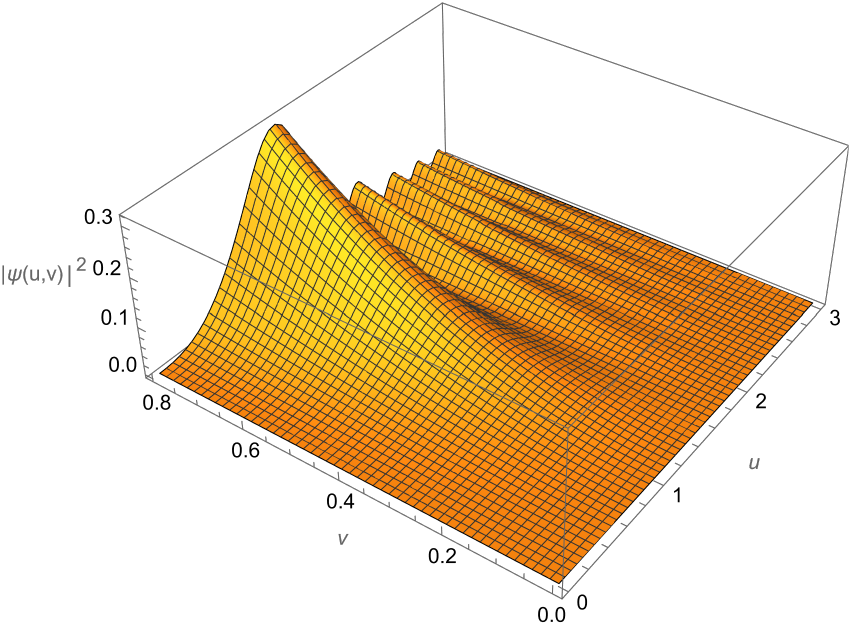}\hspace{1.9cm}
    \caption{The square of the wave function $|\psi (x,y)|^{2} $ }
  \label{fig:44}
\end{figure*}

\section{Scalar-tensor teleparallel gravity with a boundary term}

In this section, we consider the modified teleparallel model that incorporates a scalar field non-minimally coupled to both torsion and the boundary term \cite{Zubair, Wright}.

\begin{equation}\label{eq13}
S=\int \left[f(\phi)T+g(\phi)B+\frac{1}{2}\partial_{\mu}\phi\partial^{\mu}\phi-V(\phi)\right]e \ d^{4} x .
\end{equation}
In this case, $ V (\phi)$ represents the potential of the scalar field, while $ f(\phi) $ and $g(\phi)$ are both functions of the scalar field $\phi$ that exhibit smooth behavior. It is important to emphasize that this action does not introduce a new theory, but rather serves as a broader framework that encompasses and unifies various existing theories. This unification is a rolling play within the aforementioned action. To illustrate this, let us consider the following example

\begin{equation}\label{eq14}
f(\phi)= 1-\alpha\phi^{2},\ g(\phi)=-\beta\phi^{2},
\end{equation}
that $\alpha $ and $\beta$ are coupling constants. It is important to highlight that by manually selecting the values of $\alpha$ and $\beta$, we can recover scalar-tensor theories that are non-minimally coupled with the torsion scalar ($\beta = 0$), as well as theories with the boundary term $(\alpha = 0)$ and the quintessence theory $(\alpha = \beta = 0)$. The latter models have been extensively examined in the scientific literature, ranging from cosmology to astrophysical phenomena such as wormholes \cite{z114}. Additionally, the concept of traversable wormholes supported by non-minimally coupled scalar fields was initially explored in a reference \cite{z115,133}.

In this section, we take a minisuperspace FRW model with choice $ k=1$ (closed FRW), and considering the relation (\ref{eq4}), we can obtain the torsion scalar and the boundary term in the closed FRW background as follows

\begin{equation}\label{eq16}
T=\frac{6}{N^2}\left(-\frac{\dot{a}^{2}}{a^{2}}+\frac{1}{a^{2}}\right) ~~~\textrm{and}~~~~~\ B=-\frac{6}{N^2}\left(\frac{\ddot{a}}{a}+2\frac{\dot{a}^2}{a^2}-\frac{\dot{N}\dot{a}}{Na}\right).
\end{equation}

We consider the Hamiltonian constraint and its parametric solutions by substituting equations (\ref{eq15}) and (\ref{eq14}) into equation (\ref{eq13}) and introducing the variable $ \chi= \left(\frac{a \phi  }{2\sqrt{3}}\right) $. By discarding total time derivatives and integrating out the spatial degrees of freedom, we obtain the following action:

\begin{equation}\label{eq17}
S=\int dt\left(-\frac{\dot{a}^{2}a}{N}+Na+\frac{a\dot{\chi}^{2}}{N}-N\frac{\chi^{2} }{a}\right).
\end{equation}

We set $ \beta=-\frac{1}{12} $ and $\alpha=\frac{1}{12}$ to derive the above action. The point-like Lagrangian in the minisuperspace ${N,a,\chi}$ is then given by:

\begin{equation}\label{eq18}
\mathcal{L}_{\textrm{point-like}} = -\frac{\dot{a}^{2}a}{N}+Na+\frac{a\dot{\chi}^{2}}{N}-N\frac{\chi^{2} }{a}.
\end{equation}

We can derive the Hamiltonian from the Legendre transformation $C=\dot{a}P_{a}-\cal{L}$, where $P_{N}$ does not appear because it is the conjugate momentum of the non-dynamical variable $N$. The Hamiltonian is given by

\begin{equation}\label{eq19}
C=N\left[-\frac{P^{2}{a}}{4a}+\frac{P^{2}{\chi}}{4a}-a+\frac{\chi ^{2}}{a}\right]=N\cal{H},\
\end{equation}
where we have used $P_{a}=-\frac{2a\dot{a}}{N}$ and $P_{\chi}=\frac{2a\dot{\chi}}{N}$ as the canonical momenta of $a$ and $\chi$, respectively and $C$ denotes for Hamiltonian. Equation (\ref{eq19}) implies the Hamiltonian constraint or the zero energy condition $\cal{H}$$=0$, which leads to the cosmological equations. The Poisson brackets are defined as

\begin{equation}\label{eq20}
\{a, \chi\}=0, ~~~ \{P_{a},P_{\chi}\}=0, ~\{a, P_{a}\}=1~\textrm{and}~~~ \{\chi, P_{\chi}\}=1.
\end{equation}

From the Poisson brackets, we obtain the equations of motion for the scale factor and the scalar field as follows

\begin{eqnarray}\label{eq21}
\dot{a}    &=&\{a, C\}=-NP_{a}/(2a),\\\nonumber
\dot{P}_{a}&=&\{P_{a}, C\}=2N,\\\nonumber
\dot{\chi}&=& \{\chi, C\}=NP_{\chi}/(2a),\\\nonumber
\dot{P}_{\chi}&=&\{P{\chi}, C\}=-2N\chi/a.
\end{eqnarray}

If we choose the gauge $N=a$, we find the following solutions for the system:

\begin{eqnarray}\label{eq22}
a(t)&=&D\sin(t),\\\nonumber
\chi(t)&=&D\cos(t-\theta_{0}),
\end{eqnarray}

where $D$ and $\theta_{0}$ are constants. These solutions become singular at $t=0$ and $t=\pi$, which correspond to the present and future time, respectively. They are known as a Lissajous ellipsis.

\section{Quantization of the scalar-tensor teleparallel gravity and wave packets}

In this section, we will now delve into the quantum cosmology aspects of the previously presented model. Having familiarized ourselves with the approach to obtain the quantized Hamiltonian, the quantization procedure has been carried out by replacing $P_{a}$ with $-i\frac{\partial}{\partial a}$ and $P_{\chi}$ with $-i\frac{\partial}{\partial \chi}$. Consequently, the WDW is derived as ${\cal H}\psi=0$, which represents the operator version of the Hamiltonian constraint ${\cal H}=0$ and provides a description of the corresponding quantum cosmology. The Hamiltonian constraint serves as a guiding principle in this context which is given by

\begin{equation}\label{eq23}
\mathcal{H}=\frac{1}{2}\mathcal{G}_{AB}\pi^{A}\pi^{B}+\mathcal{U}(q)=0 .\\
\end{equation}
This Hamiltonian implies the WDW equation as follows,
\begin{equation}\label{eq24}
\mathcal{H}\Psi=\left(\frac{1}{2}\mathcal{G}_{AB}(-i\frac{\partial}{\partial q_{A}})(-i\frac{\partial}{\partial q_{B}})+\mathcal{U}(q)\right)\Psi=0 .\\
\end{equation}
Then by defining the Laplace-Beltrami operator as
\begin{equation}\label{eq25}
 \nabla^{2}= \frac{1}{\sqrt{-\mathcal{G}}}\partial_{A}[\sqrt{-\mathcal{G}}\ \mathcal{G}^{AB}\ \partial_{B}],\\
 \end{equation}
the final form of WDW equation reads as
\begin{equation}\label{eq26}
\left( -\frac{1}{2} \nabla^{2}+\mathcal{U}(q)\right)\Psi=0.\\
\end{equation}
where WDW metric $\mathcal{G}^{_{AB}}$ is defined by
\begin{equation}\label{eq27}
\mathcal{G}^{_{AB}}=\begin{pmatrix}
-2& 0 \\
0& 2 \\
\end{pmatrix}.
\end{equation}
Therefore, by using equations (\ref{eq24})-(\ref{eq27}), the WDW equation can be expressed as
\begin{equation}\label{eq28}
\mathcal{H}\Psi=\left\{\frac{\partial^{2}}{\partial\chi^{2}}-\frac{\partial^{2}}{\partial a^{2}}-\chi^{2}+a^{2}\right\}\Psi=0.\\
\end{equation}
We have applied a specific factor ordering and simplified the expression by absorbing a coefficient of $\sqrt{2}$ into the variables.

We can separate this partial differential equation into two ordinary differential equations in terms of the minisuperspace variables $\chi$ and $a$. The solution has the following form:
\begin{equation}\label{eq29}
	\Psi_{n}(\chi , a)=\psi_{n}(\chi)\psi_{n}(a).
\end{equation}
By inserting this expression into equation (\ref{eq28}) we have
\begin{equation}\label{eq30}
	\frac{d^{2}\psi_{n}(\chi)}{d\chi^{2}}-\chi^{2}\psi_{n}(\chi)=-E_{n}\psi_{n}(\chi),
\end{equation}
and
\begin{equation}\label{eq31}
	\frac{d^{2}\psi_{n}(a)}{d\chi^{2}}-a^{2}\psi_{n}(a)=-E_{n}\psi_{n}(a).
\end{equation}
Here, we used $E_{n}$'s as separation constants. The $\psi_{n}(z)$ is defined by following expression
\begin{equation}\label{eq32}
	\psi_{n}(z)=(\frac{1}{\pi})^{1/4}[\frac{H_{n}(z)}{\sqrt{2^{n}n!}}]e^{-z^{2}/2},
\end{equation}
where $H_{n}(z)$ is the well-known Hermite polynomial. The general form of the wave packets that satisfy the WDW equation is as follows:
\begin{equation}\label{eq33}
	\Psi(\chi,a)=\sum_{n=even}A_{n}\psi_{n}(\chi)\psi_{n}(a)+i\sum_{n=odd}B_{n}\psi_{n}(\chi)\psi_{n}(a).
\end{equation}
Moreover, to determine the solution we need the initial wave function and its derivative which are given by
\begin{equation}\label{eq34}
	\Psi_{n}(\chi,0)=\sum_{n=even}A_{n}\psi_{n}(\chi)\psi_{n}(0),
\end{equation}
\begin{equation}\label{eq35}
	\frac{\partial \Psi(\chi, a)}{\partial a}|_{a=0}=i\sum_{n=odd}B_{n}\psi_{n}(\chi)\acute{\psi_{n}}(0).
\end{equation}
It should be noted that the coefficients $A_{n}$ and $B_{n}$ play a crucial role in determining the initial wave function and its derivative, respectively. In the context of a second-order hyperbolic functional differential equation, such as equation (\ref{eq28}), these coefficients are considered to be arbitrary and independent. However, if our objective is to construct a group of wave packets that accurately simulate classical behavior, the independence of $A_{n}$'s and $B_{n}$'s is no longer maintained.

Odd functions of $a$ do not have a significant impact on the structure of the initial wave function. However, they do determine the slope of the wave function at $a=0$. Our subsequent endeavor involves examining the initial condition, thus we focus on the differential equation for small values of the scale factor. Substituting $a=0$ into the WDW equation yields the following expression.
\begin{equation}\label{eq36}
\left\{\frac{\partial^{2}}{\partial\chi^{2}}-\frac{\partial^{2}}{\partial a^{2}}-\chi^{2}\right\}\Psi=0.\\
\end{equation}
We can obtain the following differential equations by solving this partial differential equation (PDE) with the equation (\ref{eq29}).
\begin{eqnarray}\label{eq37}
\frac{d^{2}\xi_{n}(a)}{da^{2}}+E_{n}\xi_{n}(a)=0,
\end{eqnarray}
\begin{eqnarray}\label{eq38}
-\frac{d^{2}\psi_{n}(\chi)}{d\chi^{2}}+\chi^{2}\psi_{n}(\chi)=E_{n}\psi_{n}(\chi).
\end{eqnarray}

These Schr\"{o}dinger-like equations have special solutions, such as plane wave solutions for the first equation.
\begin{equation}\label{eq39}
\xi_{n}(a)=\alpha_{n}\cos \left( \sqrt{E_{n}a}\right)+i\beta_{n}\sin \left( \sqrt{E_{n}}a\right).
\end{equation}
In this expression, $ \alpha_{n} $ and $ \beta_{n} $ are arbitrary complex numbers. Equation (\ref{eq37}) is a Schr\"{o}dinger equation with the energy eigenvalue $ E_{n} $, which corresponds to the simple harmonic oscillator with the well-known solutions. The general solution of equation (\ref{eq36}) has the following form:
\begin{equation}\label{eq40}
\psi(\chi,a)=\sum_{n=even}A^{\ast}_{n}\cos( \sqrt{E_{n}}a) \psi_{n}(\chi)+i\sum_{n=odd}B^{\ast}_{n}\sin ( \sqrt{E_{n}}a)\psi_{n}(\chi).
\end{equation}
The solution above is only valid for small values of $a$. Therefore, we can derive the initial wave function and its initial slope as follows
\begin{equation}\label{eq41}
\psi(\chi,0)=\sum_{n=even}A^{\ast}_{n}\psi_{n}(\chi),
\end{equation}
\begin{equation}\label{eq42}
	\psi'(\chi,0)=i\sum_{n=odd}B^{\ast}_{n}\sqrt{E_{n}}\psi_{n}(\chi).
\end{equation}
We use the prime to indicate the derivative to the scale factor $a$. As we stated earlier, we need to find a relation between the coefficients to construct a wave packet with classical properties. We assume the coefficients have the same functional form \cite{p11,p12,bohm1, a23}.
\begin{eqnarray}\label{eq43}
A_{n}^{*}=C(n)  ~~~{\rm for} ~~n~~~{\rm even}, \\\nonumber
B_{n}^{*}=C(n)  ~~~{\rm for} ~~n~~~{\rm odd},
\end{eqnarray}
where $C(n)$ is a function of $ n $. Therefore, $A_{n}$s and $B_{n}$s are given by
\begin{eqnarray}\label{eq44}
 A_{n}=\frac{1}{\psi_{n}(0)}C(n) ~~ {\rm for} ~~~n ~~~{\rm even}, \\\nonumber
 B_{n}=\frac{\sqrt{E_{n}}}{\psi'_{n}(0)}C(n) ~~ {\rm for} ~~~ n  ~~~ {\rm odd}.
\end{eqnarray}

Note that in the above equation, without loss of generality, we can impose the conditions to our initial conditions $\psi_{n}(0)\neq0$ and $\psi'_{n}(0)\neq0$.
To construct the wave packet, we have to choose $C(n)$ so that the initial wave function matches the desired classical behavior.

By applying equations (\ref{eq33}) and (\ref{eq44}), we can obtain the precise form of the wave packet. The wave packet is given by
\begin{equation}\label{eq45}
\psi(\chi, a)=\sum_{n=even}\frac{1}{H_{n}(0)}C(n)\frac{H_{n}(\chi)H_{n}(a)}{\pi^{\frac{1}{2}}2^{n}n!}e^{\frac{-(a^{2}+\chi^{2})}{2}}+
i\sum_{n=odd}\frac{\sqrt{2n+1}}{2nH_{n-1}(0)}C(n)\frac{H_{n}(\chi)H_{n}(a)}{\pi^{\frac{1}{4}}\sqrt{2^{n}n!}}e^{\frac{-(a^{2}+\chi^{2})}{2}}.	
\end{equation}
To simplify the calculation and maintain reasonable accuracy, we used only $140$ terms in the summation instead of the infinite terms, which is sufficient and acceptable \cite{pm}.

\begin{figure*}[ht]
  \centering
  \includegraphics[width=3in]{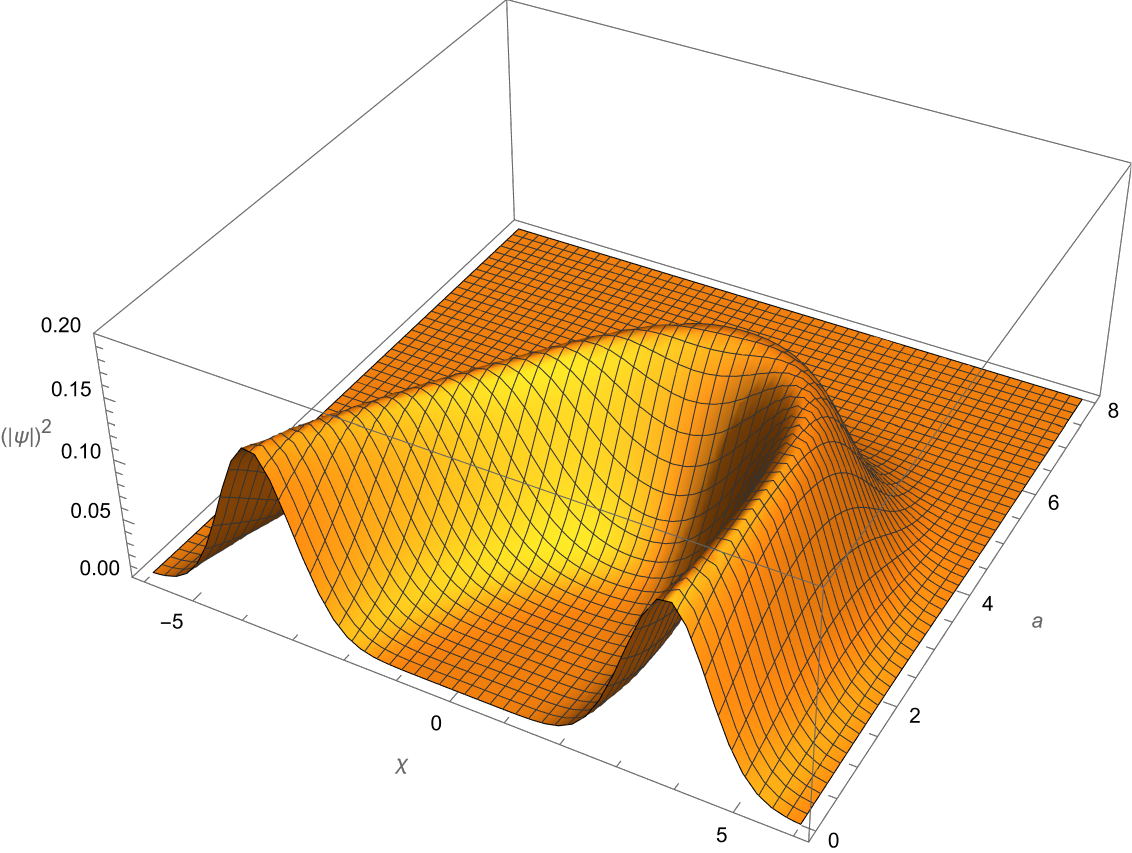}\hspace{1.9cm}
  \includegraphics[width=2in]{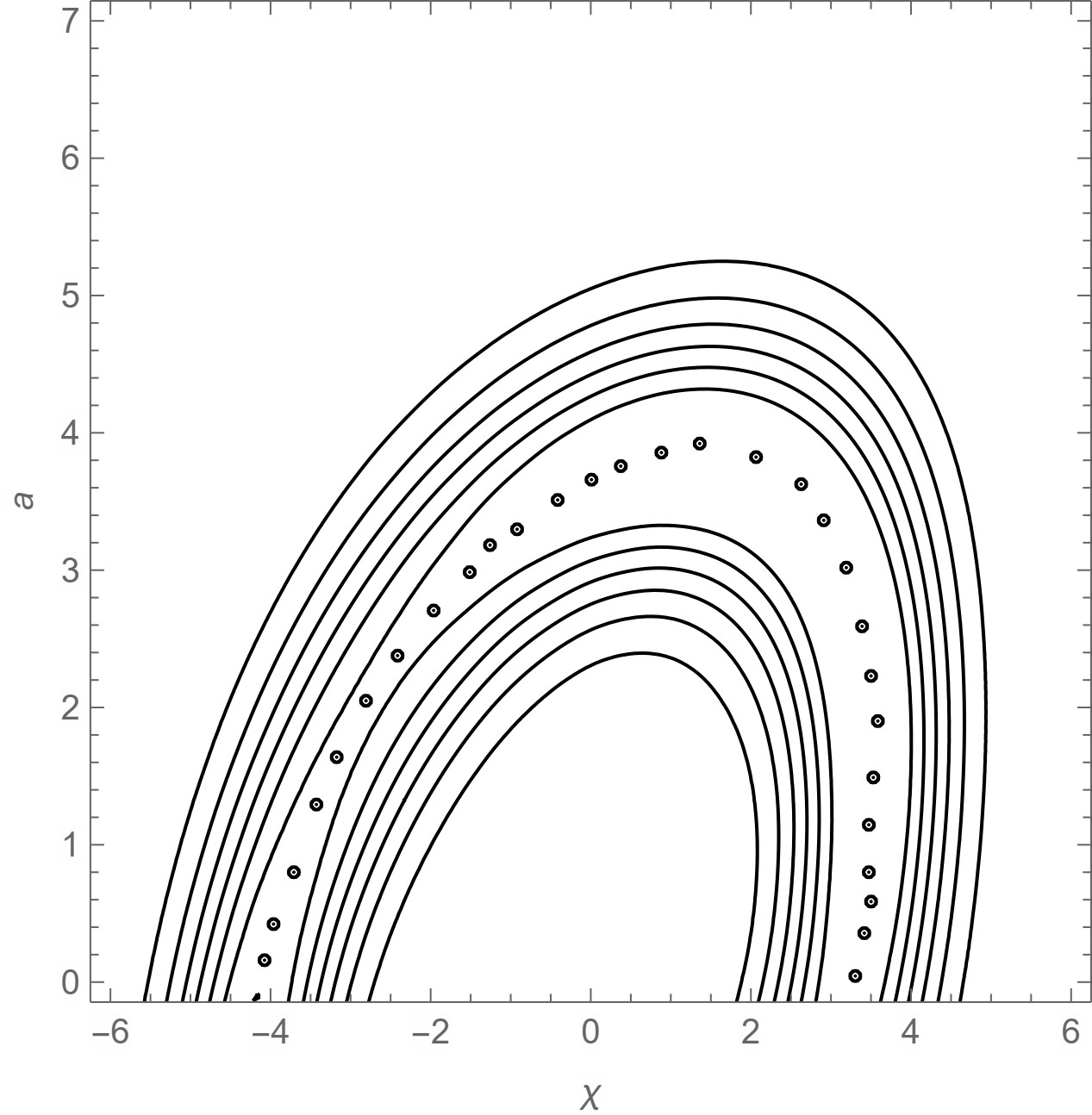}
    \caption{Left, the square of the wave packet $|\psi (\chi,a)|^{2} $ for $ C(n)=\frac{\xi^{n}}{\sqrt{2^{n}n!}}e^{\frac{-|\xi|^{2}}{4}}  $,\ $ \xi=|\xi|e^{-i\theta_{0}}  $, \ $\theta_{0}=\frac{\pi}{8}\ $ and  $  \ |\xi|=4 $. Right, the contour plot of the same figure with the classical path which inserted manually as the thick dotted line. Note that the coefficients are chosen such that the initial wave function has two well-separated peaks.}
  \label{stable}
\end{figure*}

It is important to highlight that the coefficients used to construct the diagram $\ref{stable}$ result in the initial wave function exhibiting two distinct peaks that are well separated. These peaks correspond to the classical values of $\chi$ at the beginning ($t=0$) and end ($t=\pi$) of the system. Furthermore, the diagram $\ref{stable}$ suggests the presence of a smooth wave packet with a classical behavioral crown on top of that. The figure $(\ref{stable})$ illustrates the wave packet for a specific initial condition $ C(n)=\frac{\xi^{n}}{\sqrt{2^{n}n!}}e^{\frac{-|\xi|^{2}}{4}}$, where $\xi=|\xi|e^{-i\theta_{0}}$. However, it is worth noting that it is possible to select any other suitable initial condition.

\subsection{The quantum Bohmian trajectories}

We have used the de Broglie-Bohm interpretation of quantum mechanics to study the classical and quantum correspondence. This interpretation has the advantage of showing how classical behavior emerges naturally when the quantum potential is negligible, as Bohm observed in 1952 \cite{David}.
 We can write the wave function in the polar form $ \Psi=Re^{iS} $, where $ R=R(\chi, a) $ and $ S=S(\chi, a) $ are real functions of the scale factor and the scalar field, respectively. Substituting this wave function into equation $(\ref{eq28})$, we obtain:
\begin{equation}
-\dfrac{1}{R}\dfrac{\partial^{2}R}{\partial\chi^{2}}+\dfrac{1}{R}\dfrac{\partial^{2}R}{\partial a^{2}}+(\dfrac{\partial S}{\partial \chi})^{2}-(\dfrac{\partial S}{\partial a})^{2}+\chi^{2}-a^{2}=0,	
\end{equation}
\begin{equation}
\dfrac{\partial^{2}S}{\partial\chi^{2}}-\dfrac{\partial^{2}S}{\partial a^{2}}+\dfrac{2}{R}\frac{\partial R}{\partial \chi}\frac{\partial S}{\partial \chi}-\dfrac{2}{R}\frac{\partial R}{\partial a}\frac{\partial S}{\partial a}=0.
\end{equation}
Then, by separating the real and imaginary parts of the wave packet $(\ref{eq33})$ we have
\begin{equation}
\Psi(\chi,a)=x(\chi,a)+iy(\chi,a)
\end{equation}
where, $x$ and $y$ are real functions of $ \chi $ and $a$ which are given by following equations
\begin{eqnarray}\label{ep1}
 x(\chi,a)=\sum_{n=even}\frac{1}{H_{n}(0)}Re[C(n)]\frac{H_{n}(\chi)H_{n}(a)}{\pi^{\frac{1}{2}}2^{n}n!}e^{\frac{-(a^{2}+\chi^{2})}{2}}-\sum_{n=odd}\frac{\sqrt{2n+1}}{2nH_{n-1}(0)}Im[C(n)]\frac{H_{n}(\chi)H_{n}(a)}{\pi^{\frac{1}{4}}\sqrt{2^{n}n!}}e^{\frac{-(a^{2}+\chi^{2})}{2}},
\end{eqnarray}
\begin{eqnarray}\label{ep2}
y(\chi, a)=\sum_{n=odd}\frac{\sqrt{2n+1}}{2nH_{n-1}(0)}Re[C(n)]\frac{H_{n}(\chi)H_{n}(a)}{\pi^{\frac{1}{4}}\sqrt{2^{n}n!}}e^{\frac{-(a^{2}+\chi^{2})}{2}}+\sum_{n=even}\frac{1}{H_{n}(0)}Im[C(n)]\frac{H_{n}(\chi)H_{n}(a)}{\pi^{\frac{1}{2}}2^{n}n!}e^{\frac{-(a^{2}+\chi^{2})}{2}}.
\end{eqnarray}
Then, by a straightforward calculations one can show that
\begin{equation}
R=\sqrt{x^{2}+y^{2}},	
\end{equation}
\begin{equation}
S=\arctan(\frac{y}{x}).	
\end{equation}

We can trace the Bohmian trajectories by considering the behavior of the scale factor and the scalar field, which are derived by
\begin{figure*}[ht]
  \centering
  \includegraphics[width=3in]{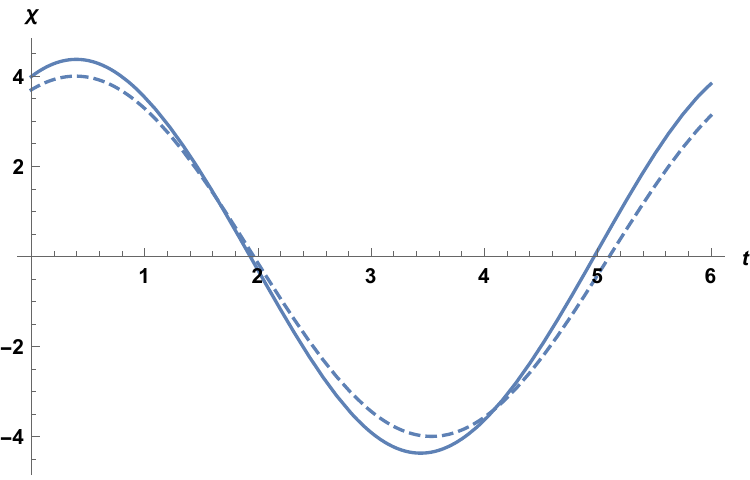}\hspace{1.9cm}
  \includegraphics[width=3in]{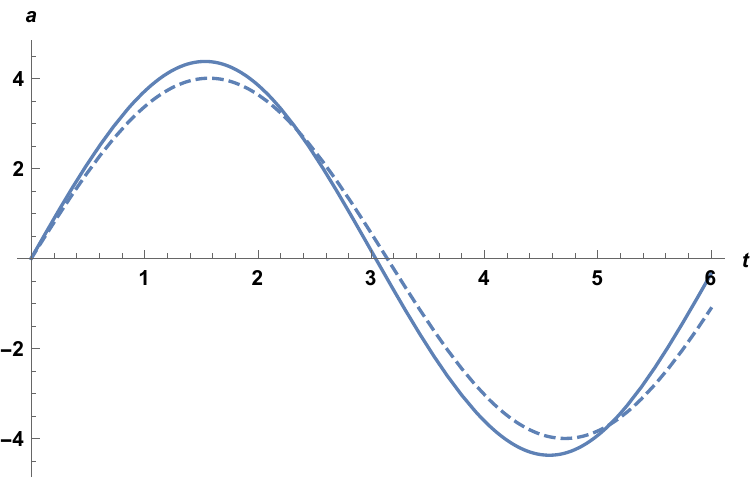}
  \includegraphics[width=4in]{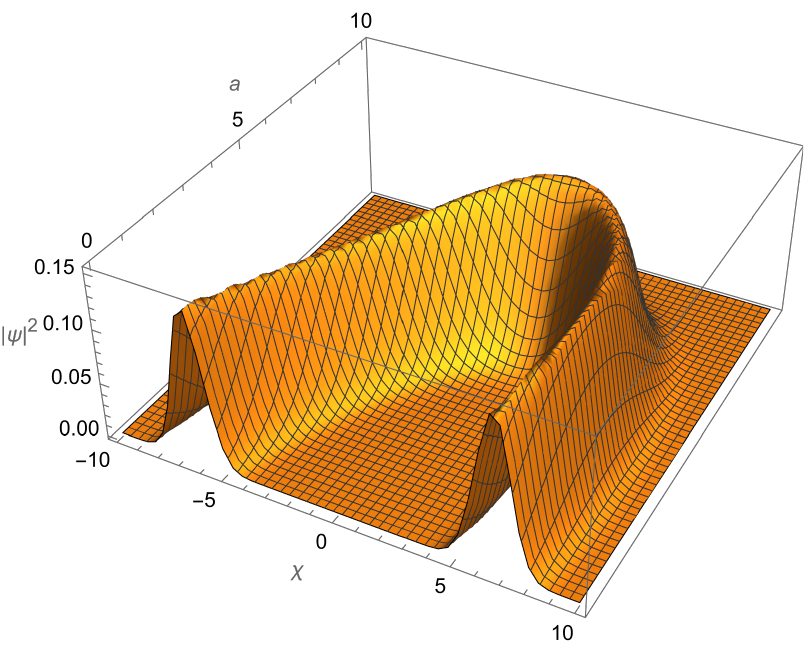}
    \caption{Up, plot of $\chi (t)$ and $a(t)$ for classical (dashed line) and Bohmian (solid line) trajectories. Down, the square of the wave packet $|\psi (\chi,a)|^{2} $ for $ C(n)=\frac{\xi^{n}}{\sqrt{2^{n}n!}}e^{\frac{-|\xi|^{2}}{4}}  $,\ $ \xi=|\xi|e^{-i\theta_{0}}  $, \ $\theta_{0}=\frac{\pi}{8}\ $ and  $  \ |\xi|=7 $.  }
  \label{fig:3}
\end{figure*}

\begin{equation}
P_{\chi}=\frac{\partial S}{\partial \chi},	
\end{equation}
\begin{equation}
P_{a}=\frac{\partial S}{\partial a}.		
\end{equation}

The canonical momenta associated with the Lagrangian (\ref{eq18}) are $P_{\chi}=2\dot{\chi}$ and  $P_{a}=-2\dot{a}$. Using the gauge choice $N=a$, we obtain
\begin{equation}
2\dot{\chi}=\dfrac{x\dfrac{\partial y}{\partial\chi}-y\dfrac{\partial x}{\partial\chi}}{x^{2}+y^{2}},
\end{equation}
\begin{equation}
2\dot{a}=-\dfrac{x\dfrac{\partial y}{\partial a}-y\dfrac{\partial x}{\partial a}}{x^{2}+y^{2}}.
\end{equation}
In the case of $ \theta_{0}\neq0$, the classical momentum is not constant and one can indicate the classical trajectory by
\begin{equation}
P=2D\sqrt{\sin^{2}(t)+\cos^{2}(t-\theta_{0})}.	
\end{equation}

The shape of the wave packet reveals the classical momentum. To see this, we apply the WKB approximation, which relates the square of the wave function to the momentum in the semiclassical limit.

\begin{equation}\label{key}
\rho=\Psi^{\ast}\Psi\propto\frac{1}{P}.
\end{equation}
One can infer from this equation that the probability density is inversely proportional to $P$, {\it i.e.}, the smaller the value of $P$, the higher the probability density, and vice versa.

\begin{figure*}[ht]
  \centering
  \includegraphics[width=3in]{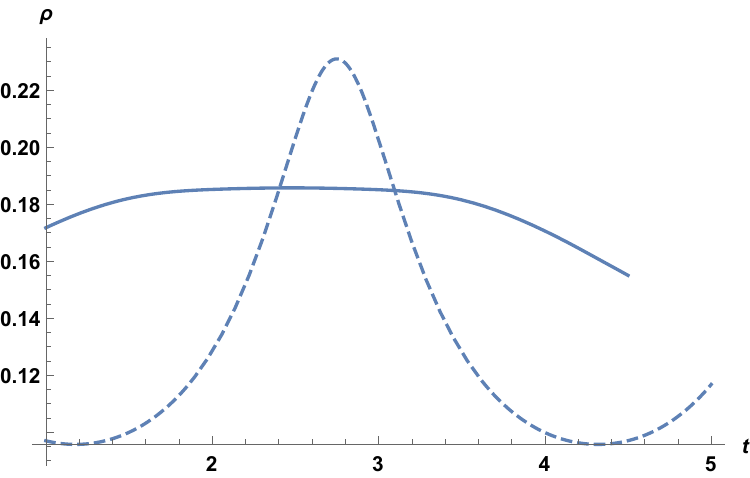}\hspace{1.9cm}
    \caption{ The inverse of classical momentum $P^{-1}$ (dashed line) and the square of the wave packet
 $|\psi (\chi, a)|^{2} $ along the classical trajectory (solid line) for  $ C(n)=\frac{\xi^{n}}{\sqrt{2^{n}n!}}e^{\frac{-|\xi|^{2}}{4}}  $,\ $ \xi=|\xi|e^{-i\theta_{0}}  $, \ $\theta_{0}=\frac{\pi}{4}\ $ and  $  \ |\xi|=4$.  }
  \label{fig:4}
\end{figure*}
By utilizing the explicit form of the wave packet equations $(\ref{ep1})$ and $(\ref{ep2})$, one can solve these differential equations numerically to determine the time evolution of $\chi$ and $a$. The lower portion of figure $(\ref{fig:3})$ illustrates the square of the wave packet $|\psi (\chi,a)|^{2}$ for a specific function $C(n)=\frac{\xi^{n}}{\sqrt{2^{n}n!}}e^{\frac{-|\xi|^{2}}{4}}$, where $\xi=|\xi|e^{-i\theta_{0}}$, $\theta_{0}=\frac{\pi}{8}$, and $|\xi|=7$. On the right side of the figure, the trajectories of $\chi(t)$ and $a(t)$ are depicted for both classical (dashed line) and Bohmian (solid line) scenarios. Remarkably, the Bohmian quantities $\chi(t)$ and $a(t)$ obtained align closely with their classical counterparts. This observation highlights the suppression of the quantum potential along the trajectory, which arises from the agreement between classical and Bohmian outcomes \cite{bohm1}.

The square of the wave packet $|\psi (\chi, a)|^{2} $ is depicted in figure $\ref{fig:4}$ along the classical (Bohmian) trajectory, which is characterized by time and the inverse of the classical momentum $P^{-1}$ versus time. From the figure, it is evident that the height of the wave packet's crest provides a qualitative representation of the variation in classical momentum along the trajectory. The lack of a strong correlation between these two quantities can be attributed to the approximate nature of equation $(\ref{key})$.

\section{Time in the quantum cosmology of scalar-tensor teleparallel gravity}

In the preceding sections, we have discussed our cosmological framework, known as scalar-tensor teleparallel gravity.  In this particular section, our objective is to preserve the cosmological constant $\Lambda$ and the presence of a free massless scalar field $\chi$. To achieve this, we introduce the cosmological variables $V:=a^{3}$, which serves as the canonical conjugate to the Hubble parameter $H$. This relationship is expressed as:
\begin{equation}
\{H, V\}=1.
\end{equation}
The momentum $p_{\chi}$ is canonically conjugated to the scalar field $\chi$. The cosmological constant $\Lambda$ is canonically conjugated to a variable denoted as ${\cal T}$, with a Poisson bracket relation $\{{\cal T},\Lambda\}=1$ \cite{bojwald}. This finding challenges the conventional understanding of the cosmological constant as a fixed value in Einstein's equation, akin to fundamental constants like $G$. However, it is mathematically consistent to treat $\Lambda$ as the momentum of the variable ${\cal T}$, even though ${\cal T}$ does not appear in the action or Hamiltonian constraint of the theory. As a consequence, the momentum $\Lambda$ remains conserved over time and manifests as a constant in the field equations. It is important to note that this alternative formulation, which introduces the canonical pair $({\cal T}, \Lambda)$, does not alter the dynamics of the theory or attempt to explain the mechanism behind dark energy. Instead, it offers a mathematically equivalent perspective that will become evident from the subsequent equations derived. Consequently, the introduction of the new parameter ${\cal T}$ provides an additional option for a global internal time, which can be compared to the more conventional global internal time $\chi$.

 For our aim, we insert cosmological constant $\Lambda$ into the constraint equation $(\ref{eq19})$  and by writing it with Hubble parameters $H$ and $N=1$ we have
\begin{equation}\label{con}
    C=-a^{3}H^{2}+\frac{P_{\chi}^{2}}{4a}-a+\frac{\chi^{2}}{a}+2a^{3}\Lambda=0.
\end{equation}
We insert the volume $V:=a^{3}$  to constraint $(\ref{con})$

\begin{equation}
    C=-VH^{2}+\frac{P_{\chi}^{2}}{4V^{1/3}}-V^{1/3}+\chi^{2}V^{-1/3}+2V\Lambda=0.
\end{equation}

We define $H$ to be canonically conjugate to volume $\{H, V\}=1$ and the cosmological constant $\Lambda$ is canonically conjugate to a variable which we call ${\cal T}$ as $\{{\cal T}, \Lambda\}=1$. The new parameter ${\cal T}$ then presents to us a new option of a global internal time.
 The proper time evolution of $V$ is
\begin{equation}
    \frac{dV}{d\tau}=\{V, C\}=2VH.
\end{equation}
Therefore, one can derive the Hubble parameter $H$ as

\begin{equation}
    H=\frac{1}{2V}\frac{dV}{d\tau}=\frac{1}{2a^{3}}3a^{2}\frac{da}{d\tau}.
\end{equation}
Therefore our proper time equation differ by a factor of $\frac{3}{2}$ with the usual $H=\frac{\dot{a}}{a}$ which implies that
\begin{equation}
    \tilde{C}=\frac{3}{2}C.
\end{equation}
We first de-parameterize the model by using the global internal times $\chi$.  We begin by solving $C=0$ for momentum $P_{\chi}$ as follows
\begin{equation}
    P_{\chi}(V,H,\Lambda,\chi)=(4V^{\frac{4}{3}}H^{2}+4V^{\frac{2}{3}}-4\chi^{2}-8V^{\frac{4}{3}}\Lambda)^{\frac{1}{2}}.
\end{equation}

In the following, we proceed with the quantization of the model after de-parameterization. This involves the introduction of an operator denoted as $p_{\chi}$, which acts upon a Hilbert space consisting of wave functions that are independent of $\chi$. An example of such a wave function is $\psi(V, {\cal T})$. For the purpose of our semiclassical analysis, we make the assumption that this operator is Weyl-ordered. By employing the techniques outlined in references \cite{d1,d2,bojwald}, we can calculate an effective Hamiltonian through a formal expansion of the expectation value as follows

\begin{eqnarray}
    H_{\chi}:=\langle P_{\chi}(\hat{V},\hat{H},\hat{\Lambda},\hat{\chi})\rangle=\langle P_{\chi}(V+(\hat{V}-V),H+(\hat{H}-H),\Lambda+(\hat{\Lambda}-H),\chi+(\hat{\chi}-\chi))\\\nonumber
    =P_{\chi}(V,H,\Lambda, \chi)+\sum_{a_{1}+a_{2}+a_{3}+a_{4}=2}^{\infty}\frac{1}{a_{1}!a_{2}!a_{3}!a_{4}!}\frac{\partial^{a_{1}+a_{2}+a_{3}+a_{4}}P_{\chi}(V,H,\Lambda, \chi)}{\partial V^{a_{1}}\partial H^{a_{2} P_{\chi}(V,H,\Lambda, \chi)}\partial \Lambda^{a_{3}}\partial \chi^{a_{4}}}\Delta (V^{a_{1}} H^{a_{2}}\Lambda^{a_{3}}\chi^{a{4}}).
\end{eqnarray}
The expansions are in $\hat{V}-V$, $\hat{H}-H$, $\hat{\Lambda}-\Lambda$ and $\hat{\chi}-\chi$. Now the symbols $V$, $H$, $\Lambda$, and $\chi$ refer to expectation values of the corresponding operators. Note that we also used them as symbols to indicate our basic variables. We have the moments $\Delta(VH\Lambda \chi)$ as independent variables and it is symmetric where for example, $\Delta(H^{2})=\Delta(H)^{2}$ is the square of the $H$-fluctuation. When the cosmological constant is considered a constant, the quantum state can be described as an eigenstate of $\Lambda$, resulting in the vanishing of all moments that involve $\Lambda$. However, despite this, we choose to retain these moments in our equations to maintain a comprehensive approach. Our analysis will focus solely on semiclassical approximations of the order $\hslash$, which encompasses corrections that are linear in second-order moments or contain terms with an explicit linear dependence on $\hslash$. Higher-order moments and products of second-order moments will be disregarded. Although the elimination of higher-order terms may not always be explicitly stated, it is applicable throughout the entirety of this paper. This approach is exemplified in our specific example. Then we have $H_{\chi}$ as equation $(\ref{hchi})$.

The expectation values and moments, viewed as functions on the space of states, are subject to a Poisson bracket induced by a commutator of operators.
These  Poisson brackets can be obtained by following the definition and  Leibniz rule.

\begin{equation}
    \{A,B\}=\frac{1}{i \hslash}\langle[\hat{A},\hat{B}]\rangle.
\end{equation}
With these definitions and  Leibniz rule we have
\begin{equation}
    \{\Delta(H^{2}),\Delta(V^{2})\}=4\Delta(VH),
\end{equation}
\begin{equation}
    \{\Delta(H^{2}),\Delta(VH)\}=2\Delta(H^{2}),
\end{equation}
\begin{equation}
    \{\Delta(V^{2}),\Delta(VH)\}=-2\Delta(V^{2}),
\end{equation}
\begin{equation}
     \{\Delta(V^{2}),\Delta(\chi H)\}=-2\Delta(V\chi),
\end{equation}
\begin{equation}
     \{\Delta(V^{2}),\Delta(\Lambda H)\}=-2\Delta(V\Lambda).
\end{equation}
Then, by following these  Poisson brackets the equation of motions are given by equations $(\ref{dv})$ and $(\ref{dh})$. Therefore, the term
\begin{equation}
    \frac{dV}{d\chi}=\{V,H_{\chi}\},
\end{equation}
is given by equation $(\ref{dv})$. And also
\begin{equation}
\frac{dH}{d\chi}=\{H,H_{\chi}\},
\end{equation}
is obtained by equation $(\ref{dh})$.

Finally, the equations of motion for the moments
\begin{equation}
  \frac{d\Delta(V^{2})}{d\chi}:=\{\Delta(V^{2}),H_{\chi}\},
\end{equation}
is given by definition $(\ref{ddelta})$. These indicate that expectation values and moments are dynamically coupled.
 In the de-parameterized setting, the absence of a quantum-corrected expression for $C$ arises from the fact that we performed the quantization of $P_{\chi}$ after resolving $C=0$. Consequently, the inclusion of proper time in a de-parameterized setting becomes ambiguous.

In the following, we introduce a new term to use a chain rule for finding the proper time equations.
\begin{equation}
    h_{\chi}:=\frac{P_{\chi}^{2}}{4V^{1/3}}+\frac{\Delta(P_{\chi}^{2})}{4V^{1/3}}+\frac{2}{9}V^{-7/3}\Delta(V^{2})-\frac{1}{6}P_{\chi}V^{-4/3}\Delta(VP_{\chi}),
\end{equation}
which leads to
\begin{equation}
    \frac{d\chi}{d\tau}=\{\chi, h_{\chi}\}=\frac{P_{\chi}}{2V^{1/3}}-\frac{1}{6}V^{-4/3}\Delta(VP_{\chi}),
\end{equation}
which one can see the full definition for it in equation $(\ref{qqq})$.

Then, by using the chain rule one can obtain the proper-time equations as follows
\begin{equation}
   \frac{dV}{d\tau}= \frac{dV}{d\chi}\frac{d \chi}{d\tau}= 2VH +\Delta(...),
\end{equation}
and
\begin{equation}
   \frac{dH}{d\tau}= \frac{dH}{d\chi}\frac{d \chi}{d\tau}=(\frac{1}{4V^{1/3}})(\frac{16}{3} H^2 \sqrt[3]{V}-\frac{32 \Lambda  \sqrt[3]{V}}{3}+\frac{8}{3 \sqrt[3]{V}})+\Delta(...).
\end{equation}
Hence, utilizing this de-parameterized approach enables us to derive the proper-time equations for the modified teleparallel model. In this model, a scalar field is non-minimally coupled to both torsion and the boundary term. It is important to mention that we did not write the term $\Delta(...)$ and their coefficients in the previous equations due to their extensive nature. However, one can easily obtain these terms by straightforwardly multiplying $\frac{dV}{d\chi}$, $\frac{d \chi}{d\tau}$, and $\frac{dH}{d\chi}$. Note that due to the length of the calculations in this section, we include some of them in Appendix A.

\section{Conclusion}

In this work, we conducted an examination of a quantum cosmology model within the framework of the modified teleparallel gravity with a boundary term. First, we considered teleparallel gravity with a general form of the boundary term and we quantized this model for the special case of $f(B)=B^{2}$. In this particular model, the wave function of the WDW equation has the ability to forecast the initial state of the universe based on its highest probability. Nevertheless, in cases where the wave function exhibits multiple peaks, it implies that various quantum states are interacting through the process of tunneling. Consequently, it suggests that our universe might have originated from diverse potential states and transitioned between different states in the past. The outcomes of this model have been summarized in figure 1, which illustrates a significant peak near certain non-zero values of $u$ and $v$, followed by smaller peaks. As the value of $u$ increases, the magnitude of these smaller peaks decreases. This indicates that the wave function has the ability to predict the initial state of the universe based on its most probable configuration. However, the presence of multiple peaks in the wave function suggests the possibility of communication between distinct quantum states through tunneling. Consequently, it implies that our universe could have evolved from various potential states and transitioned between different states in the past.

In the next section, we examine a model that includes a scalar field non-minimally coupled to both torsion and the boundary term. The wave packet of the closed FRW universe was obtained for this model, and the quantization process led to the derivation of the Hamiltonian. This Hamiltonian gave rise to the formulation of the WDW equation, which can be interpreted as an oscillator-ghost-oscillator differential equation with known solutions. The outcomes of this model are illustrated in figure 2, showing the square of the wave packet and the contour plot of the same graph. It is crucial to emphasize the reliance on the coefficients used to construct diagram 2, resulting in the initial wave function displaying two distinct, well-separated peaks. These peaks correspond to the classical values.
\\
Following the method employed to construct the wave packets, the Bohmian trajectories were determined through the de-Broglie Bohm interpretation of quantum mechanics. It is important to note that the Bohmian trajectories are significantly influenced by the wave function of the system, resulting in different trajectories based on various linear combinations of eigenfunctions. On the other hand, the underlying WDW equation represents a second-order hyperbolic functional differential equation, allowing us the flexibility to choose both the initial wave function and the initial slope of the wave function. By selecting the initial conditions thoughtfully, classical solutions were obtained. The square of the wave packet can be seen in figure 3 along the classical (Bohmian) trajectory, which is characterized by time and the inverse of the classical momentum versus time. It is clear from figure 3 that the height of the wave packet's crest offers a qualitative representation of the variation in classical momentum along the trajectory.

Ultimately, the issue of time was addressed, and through the utilization of the de-parameterization technique, which incorporates a global internal time denoted as a scalar field, the proper time equations for the scalar-tensor teleparallel gravity were derived within a semiclassical approach.

\section*{Appendix A: Some calculations related to section VI}

In this appendix, we present several enduring equations from the text.
\begin{eqnarray}\label{hchi}
H_{\chi}=(4V^{\frac{4}{3}}H^{2}+4V^{\frac{2}{3}}-4\chi^{2}-8V^{\frac{4}{3}}\Lambda)^{\frac{1}{2}}~~~~~~~~~~~~~~~~~~~~~~~~~~~~~~~~~~~~~\\\nonumber
+\frac{1}{4}(\frac{16}{9}V^{\frac{-2}{3}}H^{2}-\frac{8}{9}V^{-\frac{4}{3}}-\frac{32}{9}V^{-\frac{2}{3}}\Lambda)(4V^{\frac{4}{3}}H^{2}+4V^{\frac{2}{3}}-4\chi^{2}-8V^{\frac{4}{3}}\Lambda)^{-\frac{1}{2}}\Delta(V^{2})\\\nonumber
-\frac{1}{8}(\frac{16}{9}V^{\frac{-2}{3}}H^{2}-\frac{8}{9}V^{-\frac{4}{3}}-\frac{32}{9}V^{-\frac{2}{3}}\Lambda)^{2}(4V^{\frac{4}{3}}H^{2}+4V^{\frac{2}{3}}-4\chi^{2}-8V^{\frac{4}{3}}\Lambda)^{-\frac{3}{2}}\Delta(V^{2})\\\nonumber
+[4V^{\frac{4}{3}}(4V^{\frac{4}{3}}H^{2}+4V^{\frac{2}{3}}-4\chi^{2}-8V^{\frac{4}{3}}\Lambda)^{-\frac{1}{2}}+8V^{\frac{8}{3}}H^{2}(4V^{\frac{4}{3}}H^{2}+4V^{\frac{2}{3}}-4\chi^{2}-8V^{\frac{4}{3}}\Lambda)^{-\frac{3}{2}}]\Delta(H^{2})\\\nonumber
-[16V^{\frac{8}{3}}(4V^{\frac{4}{3}}H^{2}+4V^{\frac{2}{3}}-4\chi^{2}-8V^{\frac{4}{3}}\Lambda)^{-\frac{3}{2}}]\Delta(\Lambda^{2})\\\nonumber
-[4(4V^{\frac{4}{3}}H^{2}+4V^{\frac{2}{3}}-4\chi^{2}-8V^{\frac{4}{3}}\Lambda)^{-\frac{1}{2}}+16\chi^{2}(4V^{\frac{4}{3}}H^{2}+4V^{\frac{2}{3}}-4\chi^{2}-8V^{\frac{4}{3}}\Lambda)^{-\frac{3}{2}}]\Delta(\chi^{2})\\\nonumber
+\frac{16}{3}HV^{\frac{1}{3}}(4V^{\frac{4}{3}}H^{2}+4V^{\frac{2}{3}}-4\chi^{2}-8V^{\frac{4}{3}}\Lambda)^{-\frac{1}{2}}\Delta(VH)\\\nonumber
-2V^{\frac{4}{3}}(\frac{16}{3}V^{\frac{1}{3}}H^{2}+\frac{8}{3}V^{-\frac{1}{3}}-\frac{32}{3}V^{\frac{1}{3}}\Lambda)(4V^{\frac{4}{3}}H^{2}+4V^{\frac{2}{3}}-4\chi^{2}-8V^{\frac{4}{3}}\Lambda)^{-\frac{3}{2}}\Delta(VH)\\\nonumber
-\frac{32}{6}V^{\frac{1}{3}}(4V^{\frac{4}{3}}H^{2}+4V^{\frac{2}{3}}-4\chi^{2}-8V^{\frac{4}{3}}\Lambda)^{\frac{1}{2}}\Delta(V\Lambda)\\\nonumber
+2V^{\frac{4}{3}}(\frac{16}{3}V^{\frac{1}{3}}H^{2}+\frac{8}{3}V^{-\frac{1}{3}}-\frac{32}{3}V^{\frac{1}{3}}\Lambda)(4V^{\frac{4}{3}}H^{2}+4V^{\frac{2}{3}}-4\chi^{2}-8V^{\frac{4}{3}}\Lambda)^{-\frac{3}{2}}\Delta(V\Lambda)\\\nonumber
-4\chi(\frac{16}{3}V^{\frac{1}{3}}H^{2}+\frac{8}{3}V^{-\frac{1}{3}}-\frac{32}{3}V^{\frac{1}{3}}\Lambda)(4V^{\frac{4}{3}}H^{2}+4V^{\frac{2}{3}}-4\chi^{2}-8V^{\frac{4}{3}}\Lambda)^{-\frac{3}{2}}\Delta(V\chi)\\\nonumber
-32HV^{\frac{8}{3}}(4V^{\frac{4}{3}}H^{2}+4V^{\frac{2}{3}}-4\chi^{2}-8V^{\frac{4}{3}}\Lambda)^{-\frac{3}{2}}\Delta(H\Lambda)\\\nonumber
-32H\chi V^{\frac{4}{3}}(4V^{\frac{4}{3}}H^{2}+4V^{\frac{2}{3}}-4\chi^{2}-8V^{\frac{4}{3}}\Lambda)^{-\frac{3}{2}}\Delta(H\chi)\\\nonumber
-16\chi V^{\frac{4}{3}}(4V^{\frac{4}{3}}H^{2}+4V^{\frac{2}{3}}-4\chi^{2}-8V^{\frac{4}{3}}\Lambda)^{-\frac{3}{2}}\Delta(\chi \Lambda).
\end{eqnarray}

The term $\frac{dV}{d\chi}$ is given by

\begin{eqnarray}\label{dv}
    \frac{dV}{d\chi}=\{V,H_{\chi}\} =-\frac{4V^{\frac{4}{3}}H}{(4V^{\frac{4}
    {3}}H^{2}+4V^{\frac{2}{3}}-4\chi^{2}-8V^{\frac{4}{3}}\Lambda)^{\frac{1}
    {2}}}~~~~~~~~~~~~~~~~~~~~~~~~~~~~~~~~~~~~~~~~~~~~~~~~~~~~~~~~~~~\\\nonumber
    -[\frac{8 H}{9 V^{2/3} \sqrt{4 H^2 V^{4/3}-8 \Lambda  V^{4/3}+4 V^{2/3}-4 \chi ^2}}-\frac{H V^{4/3} \left(\frac{16 H^2}{9 V^{2/3}}-\frac{32 \Lambda }{9 V^{2/3}}-\frac{8}{9 V^{4/3}}\right)}{\left(4 H^2 V^{4/3}-8 \Lambda  V^{4/3}+4 V^{2/3}-4 \chi ^2\right)^{3/2}}]\Delta(V^{2})\\\nonumber
   +[\frac{8 H \left(\frac{16 H^2}{9 V^{2/3}}-\frac{32 \Lambda }{9 V^{2/3}}-\frac{8}{9 V^{4/3}}\right)}{9 V^{2/3} \left(4 H^2 V^{4/3}-8 \Lambda  V^{4/3}+4 V^{2/3}-4 \chi ^2\right)^{3/2}}-\frac{3 H V^{4/3} \left(\frac{16 H^2}{9 V^{2/3}}-\frac{32 \Lambda }{9 V^{2/3}}-\frac{8}{9 V^{4/3}}\right)^2}{2 \left(4 H^2 V^{4/3}-8 \Lambda  V^{4/3}+4 V^{2/3}-4 \chi ^2\right)^{5/2}}]\Delta(V^{2})\\\nonumber
\frac{16 H V^{8/3}}{\left(4 H^2 V^{4/3}-8 \Lambda  V^{4/3}+4 V^{2/3}-4 \chi ^2\right)^{3/2}}\Delta(H^{2})
-\frac{16 H V^{8/3}}{\left(4 H^2 V^{4/3}+4 V^{2/3}-8 V^{4/3}-4 \chi ^2\right)^{3/2}}\Delta(H^{2})\\\nonumber
+\frac{96 H^3 V^4}{\left(4 H^2 V^{4/3}+4 V^{2/3}-8 V^{4/3}-4 \chi ^2\right)^{5/2}}\Delta(H^{2})-\frac{192 H V^4}{\left(4 H^2 V^{4/3}-8 \Lambda  V^{4/3}+4 V^{2/3}-4 \chi ^2\right)^{5/2}}\Delta(\Lambda^{2})\\\nonumber
-[\frac{16 H V^{4/3}}{\left(4 H^2 V^{4/3}-8 \Lambda  V^{4/3}+4 V^{2/3}-4 \chi ^2\right)^{3/2}}+\frac{192 H V^{4/3} \chi ^2}{\left(4 H^2 V^{4/3}-8 \Lambda  V^{4/3}+4 V^{2/3}-4 \chi ^2\right)^{5/2}}]\Delta(\chi^{2})\\\nonumber
+[\frac{64 H^2 V^{5/3}}{3 \left(4 H^2 V^{4/3}-8 \Lambda  V^{4/3}+4 V^{2/3}-4 \chi ^2\right)^{3/2}}-\frac{16 \sqrt[3]{V}}{3 \sqrt{4 H^2 V^{4/3}-8 \Lambda  V^{4/3}+4 V^{2/3}-4 \chi ^2}}]\Delta(VH)\\\nonumber
+[\frac{64 H V^{5/3}}{3 \left(4 H^2 V^{4/3}-8 \Lambda  V^{4/3}+4 V^{2/3}-4 \chi ^2\right)^{3/2}}-\frac{24 H V^{8/3} \left(\frac{16}{3} H^2 \sqrt[3]{V}-\frac{32 \Lambda  \sqrt[3]{V}}{3}+\frac{8}{3 \sqrt[3]{V}}\right)}{\left(4 H^2 V^{4/3}-8 \Lambda  V^{4/3}+4 V^{2/3}-4 \chi ^2\right)^{5/2}}]\Delta(VH)\\\nonumber
- \frac{192 H V^{8/3} \chi }{\left(4 H^2 V^{4/3}-8 \Lambda  V^{4/3}+4 V^{2/3}-4 \chi ^2\right)^{5/2}}\Delta(\chi\Lambda)+\frac{64 H V^{5/3}}{3 \sqrt{4 H^2 V^{4/3}-8 \Lambda  V^{4/3}+4 V^{2/3}-4 \chi ^2}}\Delta(V\Lambda)\\\nonumber
-[\frac{64 H V^{4/3}}{3 \left(4 H^2 V^{4/3}-8 \Lambda  V^{4/3}+4 V^{2/3}-4 \chi ^2\right)^{3/2}}-\frac{24 H V^{7/3} \left(\frac{16}{3} H^2 \sqrt[3]{V}-\frac{32 \Lambda  \sqrt[3]{V}}{3}+\frac{8}{3 \sqrt[3]{V}}\right)}{\left(4 H^2 V^{4/3}-8 \Lambda  V^{4/3}+4 V^{2/3}-4 \chi ^2\right)^{5/2}}]\Delta(V\Lambda)\\\nonumber
+[\frac{128 H \sqrt[3]{V} \chi }{3 \left(4 H^2 V^{4/3}-8 \Lambda  V^{4/3}+4 V^{2/3}-4 \chi ^2\right)^{3/2}}-\frac{48 H V^{4/3} \chi  \left(\frac{16}{3} H^2 \sqrt[3]{V}-\frac{32 \Lambda  \sqrt[3]{V}}{3}+\frac{8}{3 \sqrt[3]{V}}\right)}{\left(4 H^2 V^{4/3}-8 \Lambda  V^{4/3}+4 V^{2/3}-4 \chi ^2\right)^{5/2}}]\Delta(V\chi)\\\nonumber
+[\frac{32 V^{8/3}}{\left(4 H^2 V^{4/3}-8 \Lambda  V^{4/3}+4 V^{2/3}-4 \chi ^2\right)^{3/2}}-\frac{384 H^2 V^4}{\left(4 H^2 V^{4/3}-8 \Lambda  V^{4/3}+4 V^{2/3}-4 \chi ^2\right)^{5/2}}]\Delta(H\Lambda)\\\nonumber
+[\frac{32 V^{4/3} \chi }{\left(4 H^2 V^{4/3}-8 \Lambda  V^{4/3}+4 V^{2/3}-4 \chi ^2\right)^{3/2}}-\frac{384 H^2 V^{8/3} \chi }{\left(4 H^2 V^{4/3}-8 \Lambda  V^{4/3}+4 V^{2/3}-4 \chi ^2\right)^{5/2}}]\Delta(H\chi),
\end{eqnarray}
and also $\frac{dH}{d\chi}$ is obtained by
\begin{eqnarray}\label{dh}
 \frac{dH}{d\chi}=\{H,H_{\chi}\}=\frac{\frac{16}{3} H^2 \sqrt[3]{V}-\frac{32 \Lambda  \sqrt[3]{V}}{3}+\frac{8}{3 \sqrt[3]{V}}}{2 \sqrt{4 H^2 V^{4/3}-8 \Lambda  V^{4/3}+4 V^{2/3}-4 \chi ^2}}~~~~~~~~~~~~~~~~~~~~~~~~~~~~~~~~~~~~~\\\nonumber
   +[\frac{-\frac{32 H^2}{27 V^{5/3}}+\frac{64 \Lambda }{27 V^{5/3}}+\frac{32}{27 V^{7/3}}}{4 \sqrt{4 H^2 V^{4/3}-8 \Lambda  V^{4/3}+4 V^{2/3}-4 \chi ^2}}-\frac{\left(\frac{16 H^2}{9 V^{2/3}}-\frac{32 \Lambda }{9 V^{2/3}}-\frac{8}{9 V^{4/3}}\right) \left(\frac{16}{3} H^2 \sqrt[3]{V}-\frac{32 \Lambda  \sqrt[3]{V}}{3}+\frac{8}{3 \sqrt[3]{V}}\right)}{8 \left(4 H^2 V^{4/3}-8 \Lambda  V^{4/3}+4 V^{2/3}-4 \chi ^2\right)^{3/2}}]\Delta(V^{2})~~~~~~~~~~~~~~~~~~~~~\\\nonumber
   +[\frac{3 \left(\frac{16 H^2}{9 V^{2/3}}-\frac{32 \Lambda }{9 V^{2/3}}-\frac{8}{9 V^{4/3}}\right)^2 \left(\frac{16}{3} H^2 \sqrt[3]{V}-\frac{32 \Lambda  \sqrt[3]{V}}{3}+\frac{8}{3 \sqrt[3]{V}}\right)}{16 \left(4 H^2 V^{4/3}-8 \Lambda  V^{4/3}+4 V^{2/3}-4 \chi ^2\right)^{5/2}}]\Delta(V^{2})~~~~~~~~~~~~~~~~~~~~~~~~~~~~~~~~~~~~\\\nonumber
   -[\frac{\left(-\frac{32 H^2}{27 V^{5/3}}+\frac{64 \Lambda }{27 V^{5/3}}+\frac{32}{27 V^{7/3}}\right) \left(\frac{16 H^2}{9 V^{2/3}}-\frac{32 \Lambda }{9 V^{2/3}}-\frac{8}{9 V^{4/3}}\right)}{4 \left(4 H^2 V^{4/3}-8 \Lambda  V^{4/3}+4 V^{2/3}-4 \chi ^2\right)^{3/2}}]\Delta(V^{2})~~~~~~~~~~~~~~~~~~~~~~~~~~~~~~~~~~~~\\\nonumber
   +[\frac{64 H^2 V^{5/3}}{3 \left(4 H^2 V^{4/3}+4 V^{2/3}-8 V^{4/3}-4 \chi ^2\right)^{3/2}}-\frac{12 H^2 V^{8/3} \left(\frac{16}{3} H^2 \sqrt[3]{V}-\frac{32 \sqrt[3]{V}}{3}+\frac{8}{3 \sqrt[3]{V}}\right)}{\left(4 H^2 V^{4/3}+4 V^{2/3}-8 V^{4/3}-4 \chi ^2\right)^{5/2}}]\Delta(H^{2})~~~~~~~~~~~~~~~~\\\nonumber
   +[\frac{16 \sqrt[3]{V}}{3 \sqrt{4 H^2 V^{4/3}-8 \Lambda  V^{4/3}+4 V^{2/3}-4 \chi ^2}}-\frac{2 V^{4/3} \left(\frac{16}{3} H^2 \sqrt[3]{V}-\frac{32 \Lambda  \sqrt[3]{V}}{3}+\frac{8}{3 \sqrt[3]{V}}\right)}{\left(4 H^2 V^{4/3}-8 \Lambda  V^{4/3}+4 V^{2/3}-4 \chi ^2\right)^{3/2}}]\Delta(H^{2})~~~~~~~~~~~~~~~\\\nonumber
   +[\frac{24 V^{8/3} \left(\frac{16}{3} H^2 \sqrt[3]{V}-\frac{32 \Lambda  \sqrt[3]{V}}{3}+\frac{8}{3 \sqrt[3]{V}}\right)}{\left(4 H^2 V^{4/3}-8 \Lambda  V^{4/3}+4 V^{2/3}-4 \chi ^2\right)^{5/2}}-\frac{128 V^{5/3}}{3 \left(4 H^2 V^{4/3}-8 \Lambda  V^{4/3}+4 V^{2/3}-4 \chi ^2\right)^{3/2}}]\Delta(\Lambda^{2})~~~~~~~~~~~~~~~~~\\\nonumber
   +[\frac{24 \chi ^2 \left(\frac{16}{3} H^2 \sqrt[3]{V}-\frac{32 \Lambda  \sqrt[3]{V}}{3}+\frac{8}{3 \sqrt[3]{V}}\right)}{\left(4 H^2 V^{4/3}-8 \Lambda  V^{4/3}+4 V^{2/3}-4 \chi ^2\right)^{5/2}}+\frac{2 \left(\frac{16}{3} H^2 \sqrt[3]{V}-\frac{32 \Lambda  \sqrt[3]{V}}{3}+\frac{8}{3 \sqrt[3]{V}}\right)}{\left(4 H^2 V^{4/3}-8 \Lambda  V^{4/3}+4 V^{2/3}-4 \chi ^2\right)^{3/2}}]\Delta(\chi^{2})~~~~~~~~~~~~~~~~~~~\\\nonumber
 +[\frac{16 H}{9 V^{2/3} \sqrt{4 H^2 V^{4/3}-8 \Lambda  V^{4/3}+4 V^{2/3}-4 \chi ^2}}-\frac{8 H \sqrt[3]{V} \left(\frac{16}{3} H^2 \sqrt[3]{V}-\frac{32 \Lambda  \sqrt[3]{V}}{3}+\frac{8}{3 \sqrt[3]{V}}\right)}{3 \left(4 H^2 V^{4/3}-8 \Lambda  V^{4/3}+4 V^{2/3}-4 \chi ^2\right)^{3/2}}]\Delta(VH)~~~~~~~~~~~~~~~~\\\nonumber
 +[\frac{3 V^{4/3} \left(\frac{16}{3} H^2 \sqrt[3]{V}-\frac{32 \Lambda  \sqrt[3]{V}}{3}+\frac{8}{3 \sqrt[3]{V}}\right)^2}{\left(4 H^2 V^{4/3}-8 \Lambda  V^{4/3}+4 V^{2/3}-4 \chi ^2\right)^{5/2}}]\Delta(VH)~~~~~~~~~~~~~~~\\\nonumber
 -[\frac{8 \sqrt[3]{V} \left(\frac{16}{3} H^2 \sqrt[3]{V}-\frac{32 \Lambda  \sqrt[3]{V}}{3}+\frac{8}{3 \sqrt[3]{V}}\right)}{3 \left(4 H^2 V^{4/3}-8 \Lambda  V^{4/3}+4 V^{2/3}-4 \chi ^2\right)^{3/2}}+\frac{2 V^{4/3} \left(\frac{16 H^2}{9 V^{2/3}}-\frac{32 \Lambda }{9 V^{2/3}}-\frac{8}{9 V^{4/3}}\right)}{\left(4 H^2 V^{4/3}-8 \Lambda  V^{4/3}+4 V^{2/3}-4 \chi ^2\right)^{3/2}}]\Delta(VH)~~~~~~~~~~~~~~~~~~~~~~~~\\\nonumber
 -[\frac{8 \sqrt[3]{V} \left(\frac{16}{3} H^2 \sqrt[3]{V}-\frac{32 \Lambda  \sqrt[3]{V}}{3}+\frac{8}{3 \sqrt[3]{V}}\right)}{3 \sqrt{4 H^2 V^{4/3}-8 \Lambda  V^{4/3}+4 V^{2/3}-4 \chi ^2}}+\frac{16 \sqrt{4 H^2 V^{4/3}-8 \Lambda  V^{4/3}+4 V^{2/3}-4 \chi ^2}}{9 V^{2/3}}]\Delta(V\Lambda)~~~~~~~~~~~~~~~~\\\nonumber
 +[\frac{2 V^{4/3} \left(\frac{16 H^2}{9 V^{2/3}}-\frac{32 \Lambda }{9 V^{2/3}}-\frac{8}{9 V^{4/3}}\right)}{\left(4 H^2 V^{4/3}-8 \Lambda  V^{4/3}+4 V^{2/3}-4 \chi ^2\right)^{3/2}}-\frac{3 V^{4/3} \left(\frac{16}{3} H^2 \sqrt[3]{V}-\frac{32 \Lambda  \sqrt[3]{V}}{3}+\frac{8}{3 \sqrt[3]{V}}\right)^2}{\left(4 H^2 V^{4/3}-8 \Lambda  V^{4/3}+4 V^{2/3}-4 \chi ^2\right)^{5/2}}]\Delta(V\Lambda)~~~~~~~~~~~~~~~~~~\\\nonumber
 +[\frac{8 \sqrt[3]{V} \left(\frac{16}{3} H^2 \sqrt[3]{V}-\frac{32 \Lambda  \sqrt[3]{V}}{3}+\frac{8}{3 \sqrt[3]{V}}\right)}{3 \left(4 H^2 V^{4/3}-8 \Lambda  V^{4/3}+4 V^{2/3}-4 \chi ^2\right)^{3/2}}]\Delta(V\Lambda)~~~~~~~~~~~~~~~~~~~~~~~~~~~~~~~~~~~~~~~~~~~~~~~~~~~~~\\\nonumber
 +[\frac{6 \chi  \left(\frac{16}{3} H^2 \sqrt[3]{V}-\frac{32 \Lambda  \sqrt[3]{V}}{3}+\frac{8}{3 \sqrt[3]{V}}\right)^2}{\left(4 H^2 V^{4/3}-8 \Lambda  V^{4/3}+4 V^{2/3}-4 \chi ^2\right)^{5/2}}-\frac{4 \chi  \left(\frac{16 H^2}{9 V^{2/3}}-\frac{32 \Lambda }{9 V^{2/3}}-\frac{8}{9 V^{4/3}}\right)}{\left(4 H^2 V^{4/3}-8 \Lambda  V^{4/3}+4 V^{2/3}-4 \chi ^2\right)^{3/2}}]\Delta(V\chi)~~~~~~~~~~~~~~~~~~~~~~~~~~\\\nonumber
 +[\frac{48 H V^{8/3} \left(\frac{16}{3} H^2 \sqrt[3]{V}-\frac{32 \Lambda  \sqrt[3]{V}}{3}+\frac{8}{3 \sqrt[3]{V}}\right)}{\left(4 H^2 V^{4/3}-8 \Lambda  V^{4/3}+4 V^{2/3}-4 \chi ^2\right)^{5/2}}-\frac{256 H V^{5/3}}{3 \left(4 H^2 V^{4/3}-8 \Lambda  V^{4/3}+4 V^{2/3}-4 \chi ^2\right)^{3/2}}]\Delta(H\Lambda)~~~~~~~~~~~~~~~~~~~~\\\nonumber
 +[\frac{48 H V^{4/3} \chi  \left(\frac{16}{3} H^2 \sqrt[3]{V}-\frac{32 \Lambda  \sqrt[3]{V}}{3}+\frac{8}{3 \sqrt[3]{V}}\right)}{\left(4 H^2 V^{4/3}-8 \Lambda  V^{4/3}+4 V^{2/3}-4 \chi ^2\right)^{5/2}}-\frac{128 H \sqrt[3]{V} \chi }{3 \left(4 H^2 V^{4/3}-8 \Lambda  V^{4/3}+4 V^{2/3}-4 \chi ^2\right)^{3/2}}]\Delta(H\chi)~~~~~~~~~~~~~~~~~~~~~\\\nonumber
 +[\frac{24 V^{4/3} \chi  \left(\frac{16}{3} H^2 \sqrt[3]{V}-\frac{32 \Lambda  \sqrt[3]{V}}{3}+\frac{8}{3 \sqrt[3]{V}}\right)}{\left(4 H^2 V^{4/3}-8 \Lambda  V^{4/3}+4 V^{2/3}-4 \chi ^2\right)^{5/2}}-\frac{64 \sqrt[3]{V} \chi }{3 \left(4 H^2 V^{4/3}-8 \Lambda  V^{4/3}+4 V^{2/3}-4 \chi ^2\right)^{3/2}}]\Delta(\chi\Lambda),~~~~~~~~~~~~~~~~~~~~~~~~~~
\end{eqnarray}
and equations of motion for the moments is given by following definition
\begin{eqnarray}\label{ddelta}
    \frac{d\Delta(V^{2})}{d\chi}:=\{\Delta(V^{2}),H_{\chi}\}~~~~~~~~~~~~~~~~~~~~~~~~~~~~~~~~~~~~~~~~~~~~~~~~~~~~~\\\nonumber
    =-4[4V^{\frac{4}{3}}(4V^{\frac{4}{3}}H^{2}+4V^{\frac{2}{3}}-4\chi^{2}-8V^{\frac{4}{3}}\Lambda)^{-\frac{1}{2}}+8V^{\frac{8}{3}}H^{2}(4V^{\frac{4}{3}}H^{2}+4V^{\frac{2}{3}}-4\chi^{2}-8V^{\frac{4}{3}}\Lambda)^{-\frac{3}{2}}]\Delta(VH)~~~~~~\\\nonumber
   -\frac{32}{3}HV^{\frac{1}{3}}(4V^{\frac{4}{3}}H^{2}+4V^{\frac{2}{3}}-4\chi^{2}-8V^{\frac{4}{3}}\Lambda)^{-\frac{1}{2}}\Delta(V^{2})~~~~~~~~~~~~~~~~~~~~~~~~~~~~~~~~~~~~~~~~\\\nonumber
   +4V^{\frac{4}{3}}(\frac{16}{3}V^{\frac{1}{3}}H^{2}+\frac{8}{3}V^{-\frac{1}{3}}-\frac{32}{3}V^{\frac{1}{3}}\Lambda)(4V^{\frac{4}{3}}H^{2}+4V^{\frac{2}{3}}-4\chi^{2}-8V^{\frac{4}{3}}\Lambda)^{-\frac{3}{2}}\Delta(V^{2})~~~~~~~~~~~~\\\nonumber
   +64HV^{\frac{8}{3}}(4V^{\frac{4}{3}}H^{2}+4V^{\frac{2}{3}}-4\chi^{2}-8V^{\frac{4}{3}}\Lambda)^{-\frac{3}{2}}\Delta(V\Lambda)~~~~~~~~~~~~~~~~~~~~~~~~~~~~\\\nonumber
+64H\chi V^{\frac{4}{3}}(4V^{\frac{4}{3}}H^{2}+4V^{\frac{2}{3}}-4\chi^{2}-8V^{\frac{4}{3}}\Lambda)^{-\frac{3}{2}}\Delta(V\chi).~~~~~~~~~~~~~~~~~~~~~~~~~~~~~~~~~\\\nonumber
\end{eqnarray}

The term for $\frac{d\chi}{d\tau}$ is given by

\begin{eqnarray}\label{qqq}
\frac{d\chi}{d\tau}=\frac{1}{2V^{1/3}}[(4V^{\frac{4}{3}}H^{2}+4V^{\frac{2}{3}}-4\chi^{2}-8V^{\frac{4}{3}}\Lambda)^{\frac{1}{2}}~~~~~~~~~~~~~~~~~~~~~~~~~~~~~~~~~~~~~\\\nonumber
+\frac{1}{4}(\frac{16}{9}V^{\frac{-2}{3}}H^{2}-\frac{8}{9}V^{-\frac{4}{3}}-\frac{32}{9}V^{-\frac{2}{3}}\Lambda)(4V^{\frac{4}{3}}H^{2}+4V^{\frac{2}{3}}-4\chi^{2}-8V^{\frac{4}{3}}\Lambda)^{-\frac{1}{2}}\Delta(V^{2})\\\nonumber
-\frac{1}{8}(\frac{16}{9}V^{\frac{-2}{3}}H^{2}-\frac{8}{9}V^{-\frac{4}{3}}-\frac{32}{9}V^{-\frac{2}{3}}\Lambda)^{2}(4V^{\frac{4}{3}}H^{2}+4V^{\frac{2}{3}}-4\chi^{2}-8V^{\frac{4}{3}}\Lambda)^{-\frac{3}{2}}\Delta(V^{2})\\\nonumber
+[4V^{\frac{4}{3}}(4V^{\frac{4}{3}}H^{2}+4V^{\frac{2}{3}}-4\chi^{2}-8V^{\frac{4}{3}}\Lambda)^{-\frac{1}{2}}+8V^{\frac{8}{3}}H^{2}(4V^{\frac{4}{3}}H^{2}+4V^{\frac{2}{3}}-4\chi^{2}-8V^{\frac{4}{3}}\Lambda)^{-\frac{3}{2}}]\Delta(H^{2})\\\nonumber
-[16V^{\frac{8}{3}}(4V^{\frac{4}{3}}H^{2}+4V^{\frac{2}{3}}-4\chi^{2}-8V^{\frac{4}{3}}\Lambda)^{-\frac{3}{2}}]\Delta(\Lambda^{2})\\\nonumber
-[4(4V^{\frac{4}{3}}H^{2}+4V^{\frac{2}{3}}-4\chi^{2}-8V^{\frac{4}{3}}\Lambda)^{-\frac{1}{2}}+16\chi^{2}(4V^{\frac{4}{3}}H^{2}+4V^{\frac{2}{3}}-4\chi^{2}-8V^{\frac{4}{3}}\Lambda)^{-\frac{3}{2}}]\Delta(\chi^{2})\\\nonumber
+\frac{16}{3}HV^{\frac{1}{3}}(4V^{\frac{4}{3}}H^{2}+4V^{\frac{2}{3}}-4\chi^{2}-8V^{\frac{4}{3}}\Lambda)^{-\frac{1}{2}}\Delta(VH)\\\nonumber
-2V^{\frac{4}{3}}(\frac{16}{3}V^{\frac{1}{3}}H^{2}+\frac{8}{3}V^{-\frac{1}{3}}-\frac{32}{3}V^{\frac{1}{3}}\Lambda)(4V^{\frac{4}{3}}H^{2}+4V^{\frac{2}{3}}-4\chi^{2}-8V^{\frac{4}{3}}\Lambda)^{-\frac{3}{2}}\Delta(VH)\\\nonumber
-\frac{32}{6}V^{\frac{1}{3}}(4V^{\frac{4}{3}}H^{2}+4V^{\frac{2}{3}}-4\chi^{2}-8V^{\frac{4}{3}}\Lambda)^{\frac{1}{2}}\Delta(V\Lambda)\\\nonumber
+2V^{\frac{4}{3}}(\frac{16}{3}V^{\frac{1}{3}}H^{2}+\frac{8}{3}V^{-\frac{1}{3}}-\frac{32}{3}V^{\frac{1}{3}}\Lambda)(4V^{\frac{4}{3}}H^{2}+4V^{\frac{2}{3}}-4\chi^{2}-8V^{\frac{4}{3}}\Lambda)^{-\frac{3}{2}}\Delta(V\Lambda)\\\nonumber
-4\chi(\frac{16}{3}V^{\frac{1}{3}}H^{2}+\frac{8}{3}V^{-\frac{1}{3}}-\frac{32}{3}V^{\frac{1}{3}}\Lambda)(4V^{\frac{4}{3}}H^{2}+4V^{\frac{2}{3}}-4\chi^{2}-8V^{\frac{4}{3}}\Lambda)^{-\frac{3}{2}}\Delta(V\chi)\\\nonumber
-32HV^{\frac{8}{3}}(4V^{\frac{4}{3}}H^{2}+4V^{\frac{2}{3}}-4\chi^{2}-8V^{\frac{4}{3}}\Lambda)^{-\frac{3}{2}}\Delta(H\Lambda)\\\nonumber
-32H\chi V^{\frac{4}{3}}(4V^{\frac{4}{3}}H^{2}+4V^{\frac{2}{3}}-4\chi^{2}-8V^{\frac{4}{3}}\Lambda)^{-\frac{3}{2}}\Delta(H\chi)\\\nonumber
-16\chi V^{\frac{4}{3}}(4V^{\frac{4}{3}}H^{2}+4V^{\frac{2}{3}}-4\chi^{2}-8V^{\frac{4}{3}}\Lambda)^{-\frac{3}{2}}\Delta(\chi \Lambda)] -\frac{1}{6}V^{-4/3}\Delta(VP_{\chi}).
\end{eqnarray}

\end{document}